\documentclass[aps,prd,a4paper,reprint,nofootinbib,superscriptaddress,floatfix,preprintnumbers]{revtex4-2}

\usepackage{xurl}
\usepackage{hyperref}
\usepackage{amsmath}
\usepackage{amsfonts}
\usepackage{amssymb}
\usepackage{color}
\usepackage[utf8]{inputenc}
\usepackage{graphicx}
\usepackage{microtype}
\usepackage{siunitx}
\usepackage{soul}
\usepackage{longtable}
\usepackage{bm}
\usepackage{xspace}
\usepackage{natbib}
\usepackage[normalem]{ulem}
\usepackage[table]{xcolor}

\usepackage{lineno}

\begin{document}

\begin{flushleft}
LAPTH-026/24
\end{flushleft}

\title{Detecting dark matter sub-halos in the Galactic plane\\with the Cherenkov Telescope Array Observatory}
\date{\today}
\author{Christopher Eckner}\email{eckner@lapth.cnrs.fr}
\affiliation{LAPTh, CNRS,  USMB, F-74940 Annecy, France}
\affiliation{Univ. of Nova Gorica, Center for Astrophysics and Cosmology, Vipavska 11, SI-5000 Nova Gorica, Slovenia}
\author{Veronika Vodeb}\email{veronika.vodeb@ung.si}
\affiliation{Univ. of Nova Gorica, Center for Astrophysics and Cosmology, Vipavska 11, SI-5000 Nova Gorica, Slovenia}
\author{Tejas Satheesh}\email{tejas.astroparticlephysics@gmail.com}\email{tsatheesh@perimeterinstitute.ca}\affiliation{LAPTh, CNRS,  USMB, F-74940 Annecy, France}
\affiliation{Perimeter Institute for Theoretical Physics, Waterloo, Ontario, N2L 2Y5, Canada}
\affiliation{Department of Physics and Astronomy, York University, Toronto, Ontario M3J 1P3, Canada}
\author{Francesca Calore}\email{calore@lapth.cnrs.fr}
\affiliation{LAPTh, CNRS,  USMB, F-74940 Annecy, France}
\author{Moritz Hütten}\email{huetten@icrr.u-tokyo.ac.jp}\affiliation{ICRR, The University of Tokyo, Japan}
\author{Pierrick Martin}\email{pierrick.martin@irap.omp.eu}
\affiliation{IRAP, Universit\'e de Toulouse, CNRS, CNES, F-31028 Toulouse, France}
\author{Gabrijela Zaharijas}\email{gabrijela.zaharijas@ung.si}
\affiliation{Univ. of Nova Gorica, Center for Astrophysics and Cosmology, Vipavska 11, SI-5000 Nova Gorica, Slovenia}
\begin{abstract}

Numerous observations confirm the existence of dark matter (DM) at astrophysical and cosmological scales, yet the fundamental nature of this elusive component of our universe remains unknown. Theory and simulations of galaxy formation predict that DM should cluster on small scales in bound structures called sub-halos or DM clumps. While the most massive DM sub-halos host baryonic matter and are observed as dwarf galaxies of the Milky Way (MW), less massive, unpopulated sub-halos could be abundant in the Galaxy as well and yield high-energy gamma rays as final products of DM annihilation. Recently, it has been highlighted that the brightest halos should also have a sizeable extension in the sky. In this study, we examine the prospects offered by the Cherenkov Telescope Array Observatory (CTAO), a next-generation gamma-ray instrument, for detecting and characterizing such objects.  
Previous studies have primarily focused on high-latitude observations; here, we assess the potential impact of the CTAO's Galactic Plane Survey, which will provide unprecedentedly deep survey data for the inner five degrees of the Galactic plane. Our modeling accounts for tidal effects on the sub-halo population, examining the conditions under which DM sub-halos can be detected and distinguished from conventional astrophysical sources. We find that regions a few degrees above or below the Galactic plane offer the highest likelihood for DM sub-halo detection. For an individual sub-halo -- the brightest from among various realizations of the MW subhalo population -- we find that detection at the 5$\sigma$ level is achievable for an annihilation cross section of $\langle \sigma v \rangle \sim 3\times10^{-25}$ cm$^3$/s for TeV-scale DM annihilating into $b\bar{b}$. For a full population study, depending on the distribution and luminosity model of Galactic sub-halos, still unconstrained cross sections in the range $\langle \sigma v \rangle \sim 10^{-23}-10^{-22}$ cm$^3$/s for TeV DM candidates are necessary for the brightest sub-halos to be detected.

\widowpenalty=5000
\clubpenalty=5000

\end{abstract}

\maketitle

\section{Introduction}\label{sec:intro}

\subsection{Weak-scale particle dark matter indirect detection}\label{sec:intro-ID}

The composition of the universe remains one of the most intriguing puzzles in the realm of astrophysics and cosmology. While visible baryonic matter constitutes only a small fraction of the total matter-energy density of the universe, a significant amount ($\sim$ 25\%) of the total matter-energy density exists in the form of dark matter (DM), a non-baryonic matter component, whose nature is still unknown~\cite{Aghanim:2018eyx}. Yet, we observe that its gravitational influence shapes the universe on a variety of scales, all the way from large-scale structure formation to dynamics within galaxies~\cite{Bergstrom:2012}.

There are many different theoretical models proposing particle DM candidates, with some of the most popular candidates being the Weakly Interacting Massive Particles (WIMPs) \cite{Goodman1985, Bertone:2010tmo}. The WIMP hypothesis, in the context of the $\Lambda$-Cold DM ($\Lambda$CDM) model of cosmology, assumes that in the early thermal history of the universe, all particles, including the DM ones, co-existed in thermal equilibrium in the primordial hot plasma (also called the thermal bath). A species of particles remains in a thermal equilibrium with other particles, as long as the interaction rate of the species with other particles in the thermal bath remains larger than the expansion rate of the universe. As the expansion rate became similar to the interaction rate of that species with other particles, that species decoupled from the thermal equilibrium, and their interaction with the thermal bath eventually stopped altogether, leaving their total number of particles of the same species constant until the present time. This way, the WIMP DM abundance freezes out. From a quantum field theoretical point of view, the simplest particle models of WIMP DM may feature particles of masses up to $\sim140$ TeV \cite{Smirnov:2019ngs} (the unitarity bound). However, in more refined models allowing for Sommerfeld enhancement, resonances or bound-state formation, this upper mass bound can be pushed beyond this zero-order estimate from unitarity (see, e.g., \cite{Baldes:2017gzw}). The WIMP hypothesis entails a rather simple mechanism to generate the observed cosmological DM abundance for GeV- to TeV-scale particles with $2\rightarrow 2$ cross sections of WIMPs and Standard Model particles at the level of the electroweak force \cite{Bertone:2005pdm} (the so-called “thermal” cross section). In addition, the WIMP hypothesis allows for testing the predicted cross section with experiments that can achieve a sufficiently good sensitivity to the DM signal. 

WIMPs have been priority targets for both astrophysical and particle physics experiments in the last decade and several different approaches to the search for DM signals have been actively pursued \cite{Garrett:2011dma}. Direct searches for DM aim at detecting the scattering of DM particles on nuclei \cite{Schumann:2019}, particle accelerator experiments look for DM signatures in the collision debris \cite{Kahlhoefer:2017}, and indirect searches scrutinize astronomical and astrophysical observations for the annihilation or decay signal of DM particles in the universe \cite{Gaskins:2016}.

Particle physics models predict that DM particles annihilate or decay into SM final states. Those final states may be unstable in nature and further decay into lighter and more stable SM particles that can traverse large distances from their place of origin to the Earth. Out of the three types of potential particle messengers, namely neutrinos, charged cosmic rays, and gamma rays, the latter represent the most prominent candidate for indirect searches. They are electrically neutral and are subject to comparatively low attenuation, which allow them to travel large distances undeflected and unabsorbed, pointing back to their origin \cite{Conrad:2017}. While this is also true for neutrinos, gamma rays interact more strongly with baryonic matter enabling a larger variety of methods to detect them in experiments. So far many gamma-ray telescopes have been utilized for the search of DM signals, observing various promising DM targets.

Over the years, a wide array of promising sources of DM annihilation or decay signals have been studied, such as the Galactic center (GC) \cite{ackermann:2017fermi, abdallah:2016search, ajello:2016fermi, gomez:2013constraints, ackermann:2013search, CTA:2020qlo}, the Galactic ridge \cite{Aharonian:2006}, Milky-Way (MW) satellites like dwarf galaxies \cite{ackermann:2014dark, walker:2011dark, magic:2016limits, Alvarez:2020cmw, DiMauro:2022hue, Kerszberg:2023cup, Butter:2023piw}, the M31 galaxy \cite{ackermann:2017observations, karwin:2021dark}, galaxy clusters \cite{ackermann:2010constraints, ackermann:2015search, DiMauro:2023qat}, globular clusters \cite{Brown:2018pwq, Bartels:2018qgr, Brown:2019teh, Beck:2023zee} and dark sub-halos within the MW \cite{diemand:2007dark, anderhalden:2013hints, kuhlen:2009exploring, madau:2008dark, Calore:2019lks}, to name a few. Due to the absence of a robustly detected signal, only upper limits and constraints on the parameter space have been derived (see \cite{Cirelli:2010xx, Doro:2014, Doro:2021dzh, Cirelli:2024ssz} for comprehensive overviews). The current situation with gamma-ray observations is that for WIMP particles of masses up to a few hundreds of GeV, the most constraining limits have been obtained from observations of dwarf spheroidal galaxies (dSphs), using space-borne and ground-based instruments, such as Fermi-LAT, H.E.S.S., MAGIC, VERITAS, and HAWC, recently combined to derive joint upper limits on the DM annihilation cross section as a function of the DM particle mass \cite{Kerszberg:2023cup}. These limits are $2 - 3$ times stronger than for individual telescopes in the multi-TeV mass range, with limits for masses below $\sim 500$ GeV dominated by \emph{Fermi}-LAT, and higher mass limits above $\sim 10$ TeV dominated by ground-based experiments. For heavier WIMP DM with masses around the TeV scale, the leading constraints are derived from H.E.S.S.~observations of the GC within the innermost $\sim 3^{\circ}$ \cite{HESS:2022ygk}. For example,  assuming a peaked DM density in the GC and $\tau^+\tau^-$ annihilation final states, the thermal cross section can already be ruled out for masses around 1 TeV given the statistical and systematic uncertainties of the study.

\subsection{Dark matter sub-halos as indirect detection targets}
\label{sec:intro-subH}



Complementary targets in gamma-ray searches are dark sub-halos which represent DM overdensities predominantly devoid of any luminous matter. They are bound to their parent DM halo and have masses between $10^{-11} - 10^{-3}\, M_\odot$ and $10^{10}\, M_\odot$, depending on the particle physics model \cite{Bringmann:2009vf}.\footnote{This statement is true mostly for ``vanilla'' WIMPs where TeV scale particles would most likely produce a cut-off closer to the lower bound. As soon as more elaborate WIMP models are considered, the cut-off scale can become sizeably larger (see, e.g., \cite{vandenAarssen:2012ag, Bringmann:2016ilk})}  They are theoretically predicted by N-body and galaxy formation simulations, implementing the bottom-up hierarchical structure formation scenario expected in the concordance cosmological model \cite{diemand:2007formation, bringmann:2009particle}. In such models, a large number of sub-halos is present in the vicinity of the Solar system. Regarding a potential luminous counterpart in DM sub-halos, only sub-halos with masses above $10^{7-8} M_\odot$ are expected to capture a sizeable amount of baryonic matter that is sufficiently large to initiate star formation leading to electromagnetic radiation over a wide range of wavelengths \cite{Sawala:2014hqa, Grand:2021fpx}. The majority of the sub-halo population is too light to produce considerable astrophysical emissions. This truly dark part of the population holds an advantage in the indirect search for DM as there is no astrophysical foreground to obscure the putative DM emission. Moreover, DM-only N-body simulations predict that sub-halos contain high concentrations of DM \cite{diemand:2008clumps, springel:2008aquarius}, potentially exhibiting DM densities that are high enough to produce a bright gamma-ray emission from DM annihilation or decay within the sensitivity reach of gamma-ray instruments \cite{Sanchez-Conde:2013yxa, Moline:2016pbm, calore:2017realistic}. Thus, they represent potentially promising DM targets.
However, one of the main challenges when looking for signals from dark sub-halos is the fact that, in this type of search, we lack information on their exact distribution within the MW. Besides, the limited resolution of N-body simulations limits the precise determination of sub-halo properties (their inner structure, abundance, etc.).

Searches for dark sub-halo signatures do not have to hinge on the WIMP hypothesis. Their gravitational influence on baryonic matter already yields viable methods to look for their presence. One approach concerns galactic stellar streams, the tidal vestiges of dwarf galaxies and globular clusters. Dynamically cooled stellar streams may be perturbed by dark sub-halos leaving, e.g.,~gaps, spurs and breaks in the stream \cite{2015MNRAS.450.1136E} depending on the stream's density. A spur candidate was identified in the GD-1 stellar stream \cite{2018ApJ...863L..20P}, which was used to infer the properties of the potential perturbing object \cite{Bonaca:2018fek, Bonaca:2020psc}. In contrast, indications for a stream break joining the seemingly disconnected ATLAS and Aliqa Uma streams were identified in \cite{2021ApJ...911..149L}. Besides individual events, the cumulative effect of multiple encounters with the Galactic dark sub-halo population on stellar streams can be studied statistically to explore the sub-halo mass function down to masses around $10^{5}\,M_{\odot}$ \cite{Bovy:2016irg} leveraging future datasets of the \textit{Gaia} mission or the Vera Rubin Observatory \cite{LSSTDarkMatterGroup:2019mwo}.

A different pure gravitational search channel is looking for distortions in Einstein rings of strongly lensed objects. The idea is that the dark substructure along the line-of-sight to the lensed object induces a magnification of the image on small scales. With this technique sub-halos with masses around $10^{8-9}\,M_{\odot}$ could be identified in data of the Hubble Space Telescope, Keck adaptive optics and ALMA interferometric data \cite{2010MNRAS.408.1969V, 2012Natur.481..341V, 2014MNRAS.442.2017V, Sengul:2021lxe}. Machine learning methods have been employed in this context and it was shown that they are capable of pushing the sensitivity of this method to less massive sub-halos rendering it possible to constrain the particle nature of DM based on the inferred sub-halo mass function \cite{Brehmer:2019jyt, Wagner-Carena:2020yun, Montel:2022fhv}.

Coming back to the WIMP hypothesis, several gamma-ray searches for dark sub-halos leveraged data of current generation gamma-ray instruments, such as unidentified sources in the \textit{Fermi}-LAT catalogs \cite{coronado:2019unidentified, coronado:2019spectral, bertoni:2015examining, calore:2017realistic, schoonenberg:2016dark, berlin:2014stringent, bloom:2012search, zechlin:2012dark, Calore:2019lks, Butter:2023piw}, HAWC \cite{Abeysekara:2019substructure, harding2015dark}, and H.E.S.S. observations \cite{glawion:2019unidentified}.
While \textit{Fermi}-LAT and current ground-based gamma-ray instruments provide stringent constraints on the DM mass and annihilation cross section, no sub-halo detection has been confirmed. A future generation of ground-based gamma-ray telescopes, the Cherenkov Telescope Array Observatory (CTAO), represents a promising next step for the discovery of dark sub-halos, by widening the reachable DM parameter space and opening the possibility of setting even more stringent constraints on DM properties. The potential of the CTAO for the detection of Galactic DM sub-halos has recently been assessed for the planned extra-galactic survey. The extragalactic survey is envisaged to cover one-quarter of the sky focusing on the northern hemisphere with a mean sensitivity to point-like sources of about 6 mCrab over the entire area covered \cite{ScienceWithCTA2019}. The authors of \cite{CoronadoBlazquez:2021} looked at the prospects of serendipitous discovery of dark sub-halos at higher latitudes, where source confusion has a less pronounced impact (see also \cite{Hutten:2016jko, Calore:2019lks}).

\subsection{Motivation for a dedicated study with CTAO}\label{sec:}
CTAO will consist of more than 60 imaging atmospheric Cherenkov telescopes (IACTs) in the northern (CTAO-North) and southern (CTAO-South) hemispheres, making it the world’s largest and most sensitive high-energy gamma-ray observatory \cite{ScienceWithCTA2019}.

CTAO surveys are expected to reach two orders of magnitude deeper exposure than the currently performed surveys using IACTs, mostly thanks to the improved sensitivity and wider field of view (FoV) \cite{ScienceWithCTA2019}. Surveys are part of CTAO's Key Science Projects which form the core scientific program of CTAO. This ensures that the surveys will receive dedicated observation time early on in the observation schedule of CTAO. The CTAO is well-suited to survey the entire Galactic plane due to its large FoV (reaching $>8^\circ$ diameter at the highest energies) and an angular resolution of a few arcminutes. The Galactic Plane Survey (GPS) is planned to be performed in the energy range between 20 GeV to 300 TeV with unprecedented depth and spectral coverage, using both CTAO-North and CTAO-South arrays.\footnote{The current plan for the GPS is outlined in detail in \cite{CTAConsortium:2023tdz}, which differs from the original description of the GPS in \cite{ScienceWithCTA2019}.} 

Several surveys of the Galactic plane have been performed with IACTs in the last decades \cite{Aharonian:2002, Abeysekara:2018}, with the largest survey performed by H.E.S.S., covering the $-110^\circ < \ell < 65^\circ$ and $|b| \leq 3^\circ$ range and detecting a total of 78 sources \cite{HESS:2018pbp}. With a target (point-source) sensitivity of the GPS at the level of a few mCrab, CTAO is expected to detect a factor of 2-7 times as many astrophysical TeV sources as there are in the current catalogs \cite{CTAConsortium:2023tdz}, offering a unique opportunity to perform population studies of Galactic very high-energy gamma-ray emitters.

Past studies have suggested different strategies to look for Galactic sub-halos \cite{Hutten:2016jko,  CoronadoBlazquez:2021}. The authors of \cite{CoronadoBlazquez:2021}, as well as most of the sensitivity predictions for DM sub-halo detectability and searches for DM sub-halos in unidentified sources, assume all DM sub-halos to be point sources. Nevertheless, N-body cosmological simulations predict possible large angular sizes for some dark sub-halos, even reaching several degrees in extension depending on their mass profile and distance, exposing the need for a dedicated study of sensitivity to extended DM sub-halos. The investigation of the detection of DM sub-halos as extended sources with \textit{Fermi}-LAT has recently been performed \cite{DiMauro:2020uos}. We focus our study on the sensitivity of the CTAO to extended DM sub-halos assuming the technical specifications of the GPS, which provides a deeper exposure with respect to the planned extra-galactic survey. The targeted (point-like) sensitivity of the GPS is around 1.8 mCrab in the inner Galaxy region, which outperforms the extragalactic survey by more than a factor of three. Despite the fact that the Galactic plane contains higher levels of astrophysical background and an increased probability of source confusion, we will show that the GPS will be competitive in the search for dark sub-halos.

The assessment of the sensitivity of the planned GPS to a DM sub-halo signal presented in this paper is a proof of principle, aimed to show that the GPS is complementary to previously suggested strategies for sub-halo detection. 
In case of no DM sub-halo detection, the sensitivity of the GPS to constrain the DM annihilation cross section is comparable to that of other targets, for example, dSphs \cite{Hiroshima:2019wvj, CTAConsortium:2023nfo}. Yet, DM sub-halos are less promising targets than the signal from DM annihilation in the GC (see Sec.~\ref{sec:main-halo}). However, if the DM profile at the GC has an extended (kpc-scale) core, the sensitivity to the DM GC signal deteriorates and depending on the extension of the density core, it could become comparable to that of satellites of the MW-like Large Magellanic Cloud \cite{CherenkovTelescopeArray:2023aqu} or the stacked limits from dSphs.



The study of the GPS sensitivity to extended sources has been performed in our paper on prospects for the detection of pulsar halos \cite{Eckner:2022gil}. In this work, we adopt a similar analysis framework and follow the same approach with a focus on dark sub-halos. 

\bigskip

The paper is organized as follows: Section \ref{sec:intro} introduces the topic of indirect searches for DM and Galactic DM sub-halos. In Section \ref{sec:model}, we list and describe the assumed background components of instrumental and astrophysical nature. We also outline the derivation of our sub-halo model and templates. The subsequent Section \ref{sec:simulation-analysis} outlines the data simulation scheme and adopted statistical framework. Section \ref{sec:sensitivity_single} reports our results for a single DM sub-halo where we obtain the detection sensitivity, discuss the impact of systematic uncertainties and show how well such an object can be distinguished from astrophysical sources once detected. The following Section \ref{sec:population} sheds light on the question of how well we can access a subset of sub-halos of the full population, and what is the contribution of the population to the overall Galactic diffuse emission. Section~\ref{sec:main-halo} discusses what are the prospects for detection of the MW's main halo in the GPS. We conclude the paper in Section \ref{sec:conclusions}.

\section{Modeling the sky components}\label{sec:model}

Following our objective to estimate the sensitivity of the CTAO GPS to DM sub-halos, we first define the sky model we employ to describe the dominant gamma-ray emission components along the Galactic plane. The main contribution to the gamma-ray emission in this region comes from the Galactic diffuse emission, whose modeling we describe first. On top of the diffuse emission, we consider different benchmark sub-halo models extracted from mock populations of Galactic sub-halos.

\subsection{Large-scale diffuse backgrounds}\label{sec:gde}
Our analyzes include the large-scale interstellar emission (IE) and examine its effects on the GPS sensitivity to DM sub-halos. 
Coming from the interaction of the Galactic CR population with the interstellar medium, the IE is a significant contribution to the gamma-ray sky. It runs predominantly along the Galactic plane and has been well-mapped at the GeV energies with the \emph{Fermi} LAT (see \cite{Fermi-LAT:2012edv} for a review). As the IE extends over large angular scales, measuring its TeV-range emission using the IACTs with very limited fields of view has been challenging. Recent advances have been made in measuring the IE at high energies using water-Cherenkov detectors (MILAGRO, HAWC, and LHAASO), thanks to their large effective area, field of view, and high duty cycles.

To model this component, we use a recent study by Luque et al. \cite{Luque:2022buq}, that takes the available GeV to PeV gamma-ray data from \emph{Fermi} LAT, Tibet AS$\gamma$, LHAASO, and ARGO-YBJ, and measurements of the local charged CR performed with AMS-02, DAMPE, CALET, ATIC-2, CREAM-III, and NUCLEON. The study uses two physical frameworks to distinguish the different assumptions on diffusion coefficient variability: the so-called 'Base' scenario models the diffusion coefficient as constant throughout the Galaxy, while the '$\gamma$-optimized' scenario allows for a radial variability of the diffusion coefficient. In our analyzes, we utilize the 'Base' framework. Both frameworks offer two spectral setups, dubbed 'Min' and 'Max,' describing different behaviors in the CR proton and Helium source spectra at high energies (see \cite{Luque:2022buq} for more details), with the 'Base-Max' setup selected as our benchmark. This model predicts the largest gamma-ray emission along most parts of the Galactic plane (except for the Galactic center region). It was also the benchmark choice for \cite{Eckner:2022gil}, in which a more detailed study of the impact of these models on searches for extended sources is given.

\subsection{Gamma rays from Galactic dark matter sub-halos}
\subsubsection{Gamma-ray DM signal}
\label{sec:dmsignal}
The signal coming from DM particle annihilation within a sub-halo depends on the density and distribution of DM and the assumed particle physics model of the interaction of DM particles with the SM. The expected flux of photons coming from DM annihilation in a given energy range is therefore
\begin{equation}
\label{eq:integr-flux}
    \mathcal{F} (E_{\mathrm{min}}, E_{\mathrm{max}}) = \frac{\langle\sigma v \rangle}{8 \pi M^2_\chi}\, \mathcal{J} \int_{E_{\mathrm{min}}}^{E_{\mathrm{max}}} \frac{\mathrm{d}N_{\chi}}{\mathrm{d}E}\, \mathrm{d}E,
\end{equation}
where $\langle\sigma v \rangle$ is the thermally averaged annihilation cross section, $M_\chi$ is the mass of the DM particle, $\mathcal{J}$ is the $\mathcal{J}$-factor corresponding to the DM sub-halo spatial density $\rho_\chi$, and $\mathrm{d}N_{\chi}/\mathrm{d}E$ is the gamma-ray spectrum of a single DM annihilation process. In our calculations, we adopt the spectral model from \cite{Cirelli:2010xx}, assuming the $b \Bar{b}$ annihilation channel and a varying DM particle mass, $M_\chi$, ranging between $100$ GeV and $100$ TeV. Given the distribution of DM within the sub-halo, the $\mathcal{J}$-factor for annihilating DM is calculated by integrating the square of the spatial density, $\rho_\chi$, over a solid angle $\Delta\Omega$ and along the line of sight (l.o.s.) $l$,
\begin{equation}
    \mathcal{J} (\Delta \Omega) = \int_{0}^{\Delta\Omega} \int_{\mathrm{l.o.s.}} \mathrm{d}l\, \mathrm{d}\Omega \, \rho_\chi^2(l, \Omega).
\end{equation}
In practice, the integration over the solid angle $\Delta \Omega$ and l.o.s.~is performed numerically via the publicly available software package \texttt{CLUMPY}\footnote{The full online documentation is hosted at: \url{https://clumpy.gitlab.io/CLUMPY/v3.1.1/}} \cite{Charbonnier:2012gf, Bonnivard:2015pia, Hutten:2018aix}. To account for the angular resolution of the CTAO, which depends on the reconstructed gamma-ray energy and to some degree also on the utilized CTAO array (CTAO-North or CTAO-South), we pixelized the $\mathcal{J}$-factor skymap using a \texttt{HEALPix} map with $N_{\mathrm{side}} = 2048$, corresponding to an angular resolution of $\sim 0.0286^{\circ}$ and a value of $\Delta \Omega = 2.5\times10^{-7}$ sr. This value is in good agreement with the spatial resolution that CTAO will reach at the highest energies, $\mathcal{O}(100$ TeV), where its spatial resolution is best \cite{ScienceWithCTA2019}. Over most of CTAO's energy range, this resolution is finer than the point-spread function (PSF) size (see also Sec.~\ref{sec:simulations} for further descriptions). Note that we do not take into account any further level of substructure boosting the $J$-factor of the MW sub-halos since even sub-halos may host sub-halos themselves as indicated by the results of $N$-body simulations (see for example \cite{Moline:2016pbm} for the modeling of the boost factor). 


\subsubsection{Sub-halo modeling: Single sub-halo with optimal properties}
\label{sec:single_SH}
Most information on the properties of Galactic sub-halo populations in the literature that take into account the baryonic effects on DM sub-structure is either extracted from numerical simulations of galaxy formation \cite{dOnghia:2010substructure, zhu:2016baryonic, kelley:2019phat, calore:2017realistic} or derived as a result of analytical or semi-analytical methods \cite{green:2007mini, goerdt:2007survival, berezinsky:2008remnants, Stref:2016uzb}. In our study, we focus on a specific semi-analytical model of Galactic sub-halos by Stref and Lavalle (SL17) \cite{Stref:2016uzb}, with the goal of providing a proof-of-principle calculation of the sensitivity of CTAO's GPS to dark sub-halos.

The SL17 sub-halo population model relies on a realistic description of the MW \cite{McMillan:2016} and takes into consideration the tidal and shocking effects in the Galactic disc. All sub-halos in the SL17 model are assumed to have a Navarro-Frenk-White (NFW) density profile \cite{navarro:1996} parametrized as:
\begin{equation}\label{eq:NFW}
    \rho_{\rm SH} (r) = \frac{\rho_s}{\left(\frac{r}{r_s}\right)\, \left(1 + \frac{r}{r_s}\right)^{2}},
\end{equation}
where $r_s$ is the so-called scale radius, defining the radius at which the steep cuspy profile transitions to a more gradual decrease, and $\rho_s$ is the characteristic density that depends on the concentration of the sub-halo. 


The resilience of DM sub-halos to tidal disruption is still under debate as the effects and strength of disruption vary significantly in different studies. Recent studies suggest that dark sub-halos are relatively resilient to tidal forces, maintaining their structures and gravitational integrity even when subject to strong gravitational potentials in their surroundings, causing significant mass loss \cite{van_den_Bosch:2017, Errani:2019}. These sub-halos can possess dense, cuspy cores, which enhance their resistance to tidal disruption. On the other hand, cosmological N-body simulations suggest that dark sub-halos are efficiently disrupted in the inner region of the Galaxy, indicating that tidal forces can significantly affect the abundance and distribution of DM sub-halos \cite{diemand:2008clumps, springel:2008aquarius}.

To account for uncertainties caused by the tidal effects, we investigate two variants of the model, dubbed “SL17 fragile” and “SL17 resilient”, originally defined in \cite{stref:2019remnants}. In the SL17 fragile model, the disruption of the sub-halo occurs when $r_t < r_s$, with $r_t$ being the tidal radius, i.e., the radius beyond which the gravitational tidal forces from the surrounding matter in the Galaxy become stronger than the self-gravity of the sub-halo. For SL17 resilient, the disruption occurs instead when $r_t < 0.01\,r_s$. As the names suggest, the SL17 fragile scenario predicts efficient disruption of dark sub-halos, while the SL17 resilient case leaves many more objects resistant to tidal disruption. For each of the two SL17 scenarios, we generated 1010 realizations of the sub-halo populations.

The results of previous works \cite{Calore:2019lks, DiMauro:2020uos} suggest a correlation between the $\mathcal{J}$-factor and the sub-halo mass and $\mathcal{J}$-factor and the angular extension, respectively. Based on these results, we expect the mass of the brightest sub-halos to lie around $10^8 M_{\odot}$. Sub-halos of mass greater than $10^8 M_{\odot}$ are massive enough to host baryonic content which can contribute significantly to the gamma-ray signal \cite{Sawala:2014hqa, Grand:2021fpx}.
As a matter of fact, the terminology ``brightest'' and ``largest $\mathcal{J}$-factor'' can be used interchangeably in this context since the expected gamma-ray signal strength is directly proportional to the $\mathcal{J}$-factor (see Eq.~\ref{eq:integr-flux}).  


\begin{figure*}[t!]
\centering
\includegraphics[width=0.48\textwidth]{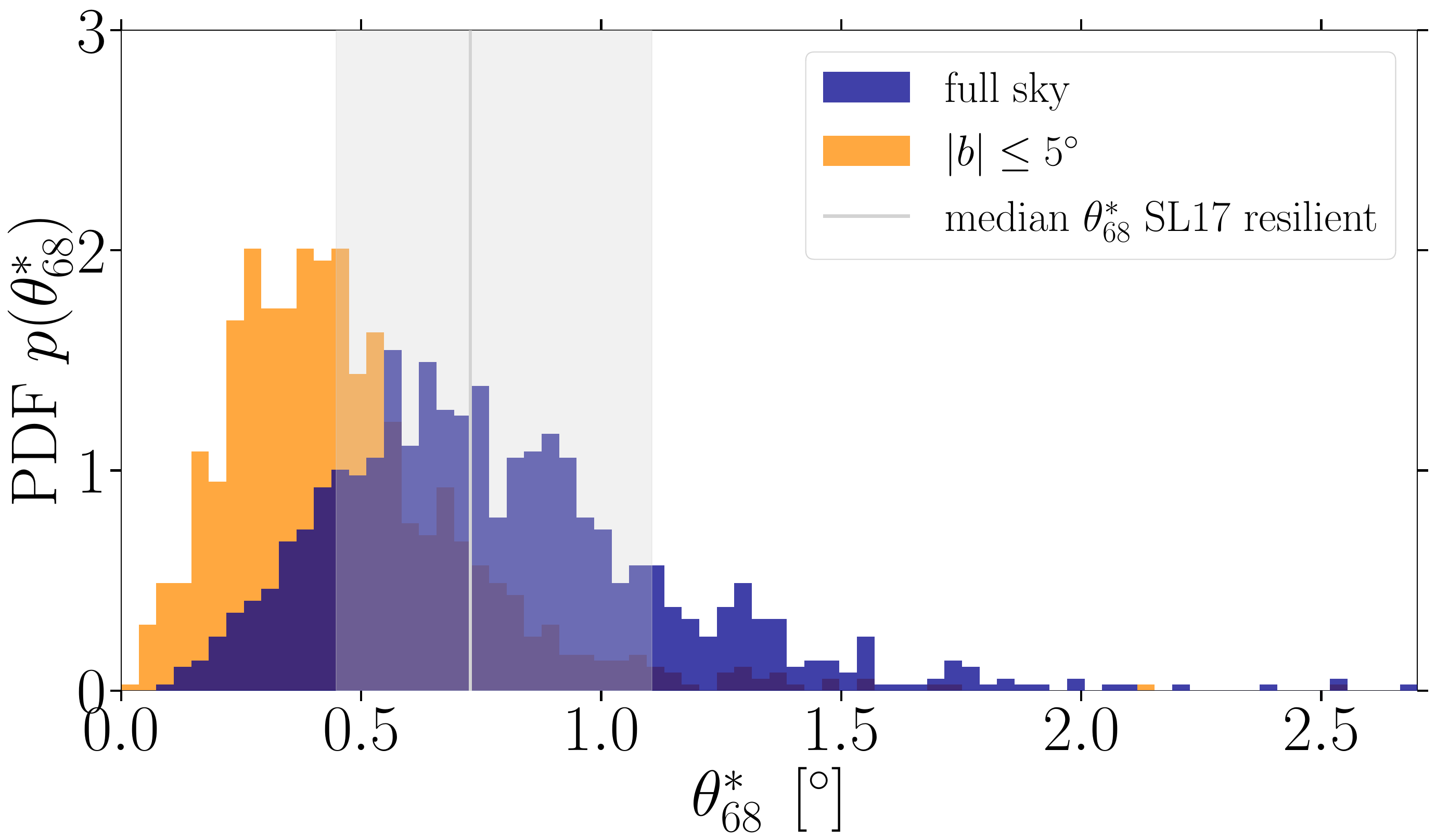}
\includegraphics[width=0.48\textwidth]{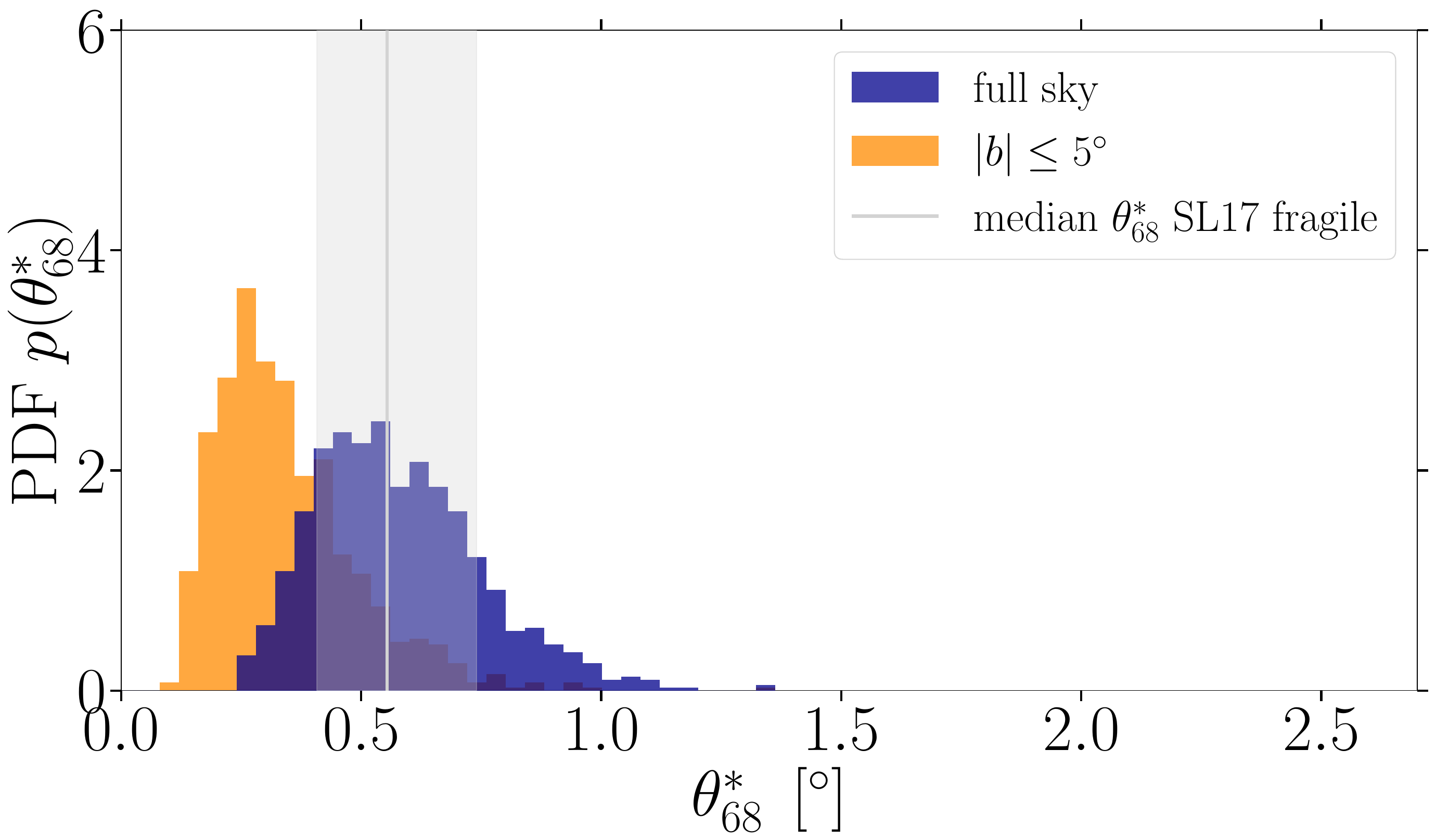}
\caption{PDF of brightest sub-halo extensions derived from all 1010 realizations of the SL17 resilient (\textit{left}) and SL-17 fragile (\textit{right}) populations (for details on the population modeling see \cite{Stref:2016uzb, stref:2019remnants}). On the $x$-axis, the angle $\theta^{\ast}_{68}$ denotes the angle in which 68\% of the total $\mathcal{J}$-factor is contained. The gray lines mark the median extension of the brightest sub-halos over all the generated populations, with the contours marking the band between the 16\% and 84\% quantiles. For comparison, we present the PDFs for selecting the brightest sub-halo from the entire sky (blue) as well as limited to the Galactic plane within $|b|\leq5^{\circ}$ (orange). Note that in contrast to \cite{DiMauro:2020uos}, the  displayed full-sky histograms only include subhalos up to $M_{\Delta} = 10^8\,M_{\odot}$.}
\label{fig:SL17_extensions}
\end{figure*}



We investigate CTAO's potential to detect a single DM sub-halo in the most optimistic scenario. The brightest sub-halos should provide the best sensitivity. To study CTAO's sensitivity to an ideal sub-halo, we focus on sub-halos up to $10^8 M_{\odot}$ with high DM density, yielding large $\mathcal{J}$-factors. By browsing all of our realizations and SL17 scenarios, we find the most optimistic sub-halo with $r_s = 2.36$ kpc, $\rho_s = 3.4\cdot 10^6 M_{\odot}/\mathrm{kpc}^3$, $M_{\Delta} = 4.8\cdot10^6\,M_{\odot}$ and $r_{\Delta} = 0.162$ kpc at a distance of $d = 0.92$ kpc. For us, the quantity $r_{\Delta}$ of a DM subhalo defines its total extent. It is the radius up to which the overall sub-halo mass equals the tidal mass $M_{\Delta}$, i.e.~the mass associated with the sub-halo after evolving it in the MW's gravitational potential and accounting for tidal stripping and shocking effects. 

Since we set out to determine CTAO's sensitivity to \textit{extended} sub-halos, we define the total angular extension of a sub-halo as $\arctan(r_{\Delta}/d)$. The extension depends, on one hand, on the evolution of the sub-halo within the MW since it determines $r_{\Delta}$ and, on the other hand, on the distance of the sub-halo $d$ to the observer. A second useful quantity providing information about the sub-halo's extent is $\theta_{68}$ defined as the radius or angular size that contains 68\% of the sub-halo's total $\mathcal{J}$-factor. We adopt $\theta_{68}$ as the benchmark quantity when referring to the angular size of a sub-halo following the approach in \cite{DiMauro:2020uos}.

In our search for the most optimistic sub-halo, we browsed the available set of population realizations for the sub-halo with the largest $\mathcal{J}$-factor. We observe a certain scatter of the associated angular extent for both SL17 scenarios. In Fig.~\ref{fig:SL17_extensions}, we show the Probability Distribution Function (PDF) for the sample of $\theta_{68}$ obtained from the brightest sub-halo of each realization, i.e.~$\theta^{\ast}_{68}$, searching in the full sky (blue) and only in the Galactic plane at $|b|\leq5^{\circ}$ (orange). Focusing first on the distributions we derive for the entire sky, the brightest sub-halos of the resilient scenario exhibit generally larger extensions than their counterparts in the fragile scenario. This observation is caused by the fact that resilient sub-halos are on average closer to Earth than sub-halos of the fragile scenario since the former may survive the gravitational forces present in the inner parts of the MW. Therefore, their apparent size in the sky is larger while their overall size $r_{\Delta}$ is typically smaller than found in fragile populations. Moreover, the PDF for the SL17 resilient model exhibits a substantial tail with extensions up to $2.5^{\circ}$ that is absent in the SL17 fragile case. This tail is, again, caused by the ability of resilient sub-halos to survive as bound objects at smaller distances from the GC than fragile sub-halos. 
In comparison, the brightest sub-halos specifically localized in the Galactic plane region show a smaller extension on average. This observation is in line with the intuition that the fraction of surviving sub-halos was more severely affected by the baryonic gravitational potential of the Galactic disc; thus, inducing stronger tidal stripping or shocking effects. The tail of the PDFs can reach extensions similar to the maximal extensions over the entire sky, though. These findings also imply that, on average, dark sub-halos in the Galactic plane will appear more point-like than sub-halos found at higher latitudes. Therefore, the GPS can naturally complement\footnote{Of course, this statement depends on the observer's distance to the bright part of the Galactic dark sub-halos population.} the search for Galactic sub-halos in data of CTAO's extragalactic survey and provide competitive constraints as we will derive in what follows. 

Additionally, we display in Fig.~\ref{fig:SL17_extensions} the median extension of the brightest sub-halos over the entire sky as a vertical gray line while the 16\% and 84\% quantiles are visualized as gray bands. We later use sub-halo templates representing these median cases as references to infer the sensitivity to the SL17 resilient and SL17 fragile populations as a whole as outlined in Sec.~\ref{sec:details-integrated-sensitivity}. We chose the full-sky samples and their median for this study because we aim to gauge our method against larger extensions which are still plausible even for sub-halos in the Galactic plane given the wide tails of the orange distributions. 

\section{Data simulation and analysis}\label{sec:simulation-analysis}

\subsection{Simulation of GPS observations}\label{sec:simulations}

We adopt the approach described in \cite{CTAConsortium:2023tdz} to simulate the GPS observations. The planned observation schedule consists of an early-phase program (480 hours of observation time in the first two years of observations) and a long-term program (1140 hours of observation time over the following eight years). The planned pointing strategy for the GPS adopts a double-row pattern of equally spaced pointing directions, observations during the dark time (moonless nights), and optimizations regarding minimal zenith angles of each observation. The exposure of different regions of the Galactic plane will differ depending on the expected number of sources and objects of interest in each region. The Galactic plane is divided into five respective regions, namely the Inner Galaxy, which is planned to receive more observation time and with that deeper exposure, as it is most densely populated with sources; the Cygnus/Perseus region; the Vela/Carina region; the outer South region; and the Anticenter region. Therefore, each region's point- and extended-source sensitivity will differ, which is considered when performing our population study.

To simulate GPS observations, we utilize the most recent Instrument Response Functions (IRF) provided by the CTAO\footnote{See \url{https://www.ctao.org/for-scientists/performance/} for more details.}, dubbed \texttt{prod5-v0.1}\footnote{The IRF files are publicly available at \cite{cherenkov_telescope_array_observatory_2021_5499840}.}. The IRFs correspond to the anticipated configuration of the CTAO arrays, consisting of 4 Large- and 9 Middle-Sized Telescopes (STs) in the Northern site and 14 Middle- and 37 Small-STs in the Southern site. These represent the instruments that will be built in the initial construction phase. Among other properties, the IRFs contain information about CTAO's PSF and energy dispersion. 
Simulations and analyzes were performed using the publicly available gamma-ray analysis software \texttt{gammapy}\footnote{\url{https://gammapy.org}} \cite{gammapy:2023, gammapy:zenodo-1.2}.
We selected the set of IRFs corresponding to background cuts appropriate for studies of extended sources. These cuts were optimized based on Monte Carlo simulations of 50 h of observations. The largest fraction of events detected by ground-based observatories come from air showers triggered by cosmic-ray (CR) particles that enter the atmosphere. These can be misidentified as gamma rays in the event reconstruction process and represent the irreducible CR background component. This CR component is modeled from extensive Monte-Carlo simulations of CR air showers, detection of the associated Cherenkov radiation, and subsequent event reconstruction. The rate and distribution of misidentified CR events are provided by the IRF suite and depend on the given observing conditions, such as the duration of observations and the zenith angle. This component is fitted to the data, together with the astrophysical emission models (cf.~Sec.~\ref{sec:model}), in the case of a real data analysis. In the spatial-spectral likelihood analysis introduced in the next section, the overall normalization of the CR background is left free to account for the uncertainties or biases in this component.

\subsection{Statistical analysis framework}\label{sec:statistics}


We adopt the statistical inference framework of Ref.~\cite{CTA:2020qlo} and adapt it to our needs, as it was done in our previous work on detecting and characterizing pulsar halos \cite{Eckner:2022gil}. Consequently, we are conducting binned template-based fits of gamma-ray emission models to simulated CTAO mock data. Given a specific region of interest, we devise a gamma-ray emission model as a linear combination of templates encompassing the expected CR background $B$, taken from the adopted CTAO IRF instance, and astrophysical signal components $\{S_k\}_{k\in K}$. The latter set of templates $\{S_k\}_{k\in K}$ is comprised of components describing conventional astrophysical gamma-ray emission like the diffuse emission originating in CR interactions with the interstellar medium as well as exotic contributions linked to DM annihilation in sub-halos of the MW's parent halo. All statistical inference is based on the Poisson likelihood function:
\begin{equation}
\mathcal{L}\!\left(\left.\bm{\mu}\right|\bm{n}\right)=\prod_{i,j} \frac{\mu_{ij}^{n_{ij}}}{\left(n_{ij}\right)!}e^{-\mu_{ij}}\mathrm{,}
\end{equation}
where $\bm{\mu}$ represents the gamma-ray emission model (cf.~Sec.~\ref{sec:model}) used to fit $\bm{n}$, the CTAO experimental mock data. The index $i$ enumerates the energy bins while the index $j$ labels the spatial pixels of the employed templates.

Our gamma-ray emission model is then generally given by:
\begin{equation}
\bm{\mu}= \sum_{k\in K} \bm{S}_k(\bm{\theta}^S_k) + \bm{B}(\bm{\theta}^B) \mathrm{,}
\end{equation}
where $\{\bm{\theta}^S_k\}_{k \in K}$ and $\{\bm{\theta}^B\}$ represent generic model parameters adjusting the spectral and angular dependence of the signal and CR background templates. For all practical matters, we consider linear models assigning renormalization parameters (global or per energy bin) to our employed templates, i.e.~spelt out explicitly
\begin{equation}
\label{eq:model_spe_eq}
\mu_{ij}= \sum_{k\in K} \theta^S_{k, i} S_{k,ij} +  \theta_i^B B_{ij} \mathrm{.}
\end{equation}

We derive CTAO sensitivity projections based on the log-likelihood ratio test statistic (TS). This approach allows us to statistically quantify the evidence for detecting a new signal above a given background or to distinguish between alternative hypotheses regarding the nature of a measured signal. The TS is generically expressed as:
\begin{equation}
\label{eq:tsdet}
\textrm{TS} = 2 \left( \ln \left[ \frac{\mathcal{L} \left( \bm{\mu} ( { \hat{\bm{\theta}}_k^{S_{\rm test}} }, { \hat{\bm{\theta}}^B }) | \bm{n} \right)}{\mathcal{L} \left( \bm{\mu} ({ \hat{\bm{\theta}}^{S_{\rm null}}_k }, { \hat{\bm{\theta}}^B}) | \bm{n} \right)} \right] \right) \mathrm{,}
\end{equation}
where the hatted quantities refer to the best-fitting values for all model parameters. The model composition in the null and test hypotheses depends on the task at hand. It will be specified for each analysis individually.

\subsection{General analysis definitions}
\label{sec:single-halo-analysis-details}

In the case, we analyze single sub-halos (cf.~Sec.~\ref{sec:single_SH} for the description of the brightest sub-halo candidate), the template-based likelihood analysis is performed using the following specifications:
\begin{itemize}
    \item The sub-halo is positioned at the reference location $(l, b) = (40^\circ, 0^\circ)$ of our region of interest (ROI), represented by $6^\circ \times 6^\circ$ regions around the sub-halo positions. We will use this location in the Inner Galaxy region of the GPS, which is characterized by the highest exposure (an illustration of the exposure across the GPS band can be found in \cite{Eckner:2022gil}). This position is far from the rather complex GC, which allows us to reduce the amount of background component modeling for this sensitivity forecast.
    \item Simulations are performed on a $0.02^\circ \times 0.02^\circ$ pixel grid unless stated otherwise. The simulated templates are later re-binned to a larger, $0.12^\circ \times 0.12^\circ$ grid before performing the sensitivity analysis to ensure that the chosen bin size corresponds to the characteristic length scale of CTAO PSF (see Sec.~\ref{sec:sens_iemsyst} for more details).\footnote{The spatial bin size for the sensitivity analysis does not strongly impact the forecast when no systematic uncertainties are included. Otherwise, our treatment of systematic uncertainties introduces a dependence on the spatial bin size and, thus, our results should be understood in this sense. A more detailed study of the impact of the spatial bin size on sensitivity forecasts with CTAO is given in \cite{CTA:2020qlo} in the context of searches for DM annihilation from the Milky Way main halo in the Galactic center.}
    \item To properly sample the IRFs and in particular the energy dispersion of the array, the energy binning is done with 15 logarithmically spaced energy bins per decade. The energy range we consider is between 10 GeV and 100 TeV. The finer energy bins are re-binned to 5 log-spaced energy bins per decade before the likelihood analysis. However, not all energy bins will contain statistically significant numbers of events (at lower energies due to exceeding the low limit of the CTAO's energy range and at high energies due to a DM mass cutoff in the spectra). Those energy bins will not be considered in the spatial-spectra likelihood analysis which is predominantly done bin-by-bin, as we will detail in Sec.~\ref{sec:source-discrimination}. 
\end{itemize}

Considering single DM sub-halos, we will perform several analyzes that apply Eq.~\ref{eq:tsdet}. In what follows, we provide a brief description of how this equation is specified and approached in each of these cases.

\subsection{Deriving the differential flux sensitivity to single sub-halos} 

This analysis part revolves around the most optimistic scenario for a dark sub-halo detection based on the brightest sub-halo candidate identified in our population realizations. However, the selected optimal DM sub-halo yields a single sensitivity estimate for CTAO that does not capture the dependence on all the parameters that enter the value of $\theta_{68}$. To study the CTAO sensitivity to extended sub-halos in a broader sense, we employ different approaches:
\begin{itemize}
    \item First, we vary the distance of the optimal sub-halo keeping the radius $r_{\Delta}$ constant. Besides its nominal distance $d\sim1$ kpc, we simulate the sub-halo for three additional distances $d = 5, 10, 30$ kpc.
    \item Secondly,  we vary the angular extension $\theta_{68}$ by keeping the distance constant but selecting different brightest sub-halos from the realizations that are nominally found at $d=1$ kpc in their respective population realization. We consider the alternative cases of $\theta_{68} = 1.24^{\circ}, 0.58^{\circ}$.
\end{itemize}

The sensitivity of CTAO depends on the location of the sub-halo as well since the GPS consists of regions with varying exposure as mentioned in Sec.~\ref{sec:simulations}. Since we find sub-halos all along the Galactic plane in our simulations, we consider the single benchmark location given in the previous section, at which we place the center of our selected DM sub-halo templates. The sensitivity to sub-halos in the other regions of the GPS can easily be derived by re-scaling the sensitivity we obtain for the Inner Galaxy. Since the pointing pattern of the GPS is designed to achieve a relatively uniform exposure in the different GPS domains, the difference in exposure is typically a factor of a few following from the total observation time $t$ so that the sensitivity will scale with $\sqrt{t}$.

A list of all considered single sub-halo benchmarks and their properties are given in Tab. \ref{tab:sub-halos}. The obtained results for the sensitivity to the different sub-halos described above and listed in the table are discussed in Sec.~\ref{sec:sensitivity_single}. The differential sensitivity and the constraining power to the DM annihilation cross section for different sub-halos are displayed in Sec.~\ref{sec:flux-sensitivity}.

Having defined the cases to consider for this analysis, we derive the GPS $\mathrm{TS}=25$ ($5\sigma$) detection sensitivity per energy bin stated in units of flux (differential flux sensitivity) in complete analogy to \cite{Eckner:2022gil} with the only difference being the employed spatial template, i.e.~the respective DM sub-halo profile. For details, for instance, how we effectively include systematic uncertainties, we refer the reader to this publication.

\subsection{Morphological discrimination of dark sub-halos}

\begin{figure}
\centering
\includegraphics[width=\columnwidth]{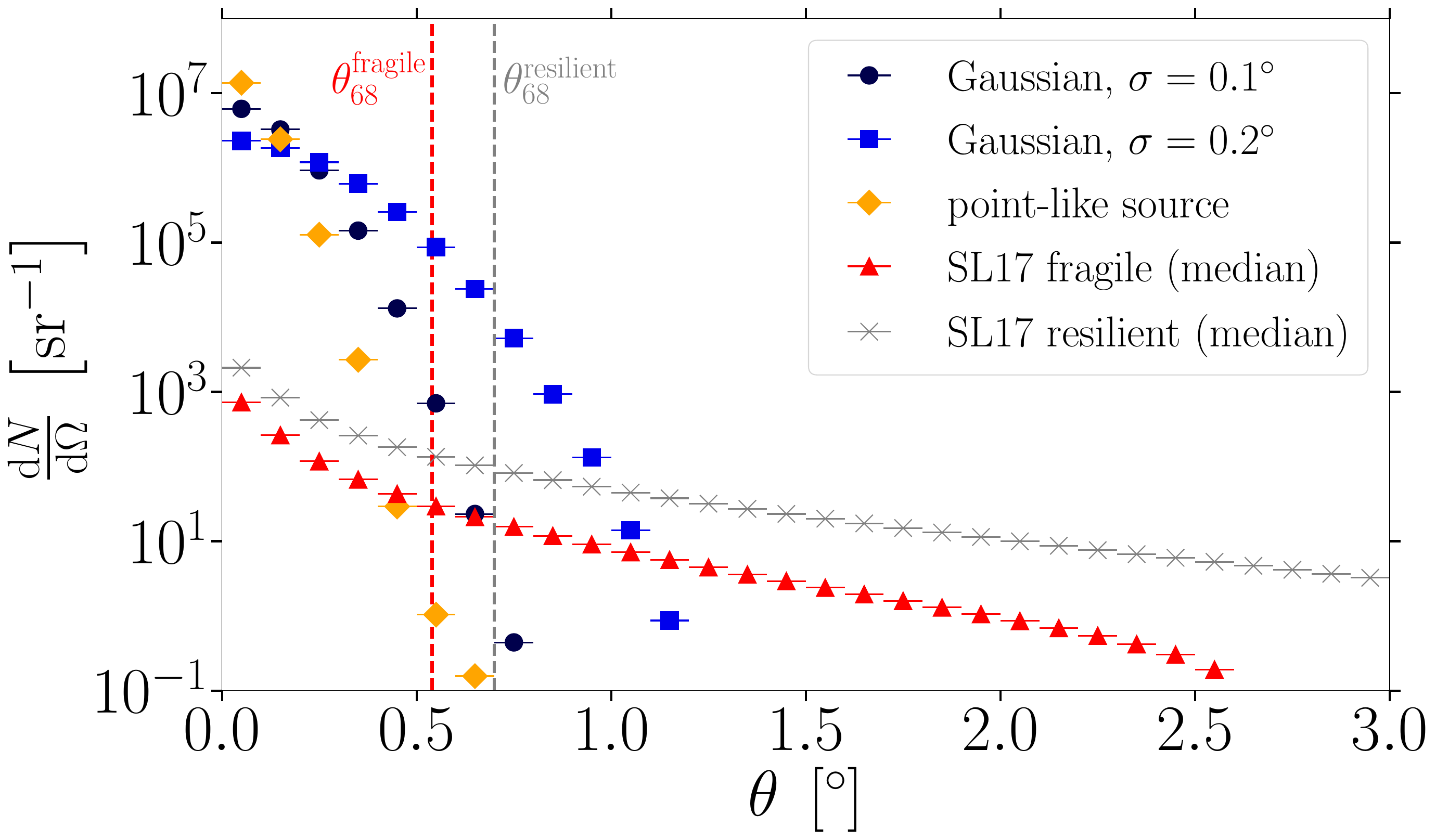}
\caption{Comparison of the surface brightness angular profile (concentric annuli of width $0.1^{\circ}$) of two DM sub-halo templates, SL17 resilient (gray) and fragile (red), corresponding to the median angular extension marked with a vertical gray line in Fig.~\ref{fig:SL17_extensions}. The angular extension $\theta_{68}$ of SL17 fragile amounts to $0.54^{\circ}$ while the resilient sub-halo exhibits an extension of $0.7^{\circ}$. Both angular sizes are marked with vertical dashed lines in their respective color. For reference, we show the surface brightness profile of a point-like source and two extended sources following a Gaussian profile of a width of $0.1^{\circ}$ and $0.2^{\circ}$, respectively. The profiles are convolved with the CTAO IRFs and extracted from a simulation between 100 GeV to 1 TeV. We assume $M_{\chi} = 100$ TeV and annihilation into $b\bar{b}$ final states.}
\label{fig:DMsubhalo-extension}
\end{figure}

The NFW density profile that characterizes each DM sub-halo leads to a very particular spatial morphology that distinguishes them from more conventional astrophysical objects. We compare the appearance of DM sub-halos to typical profiles encountered for astrophysical sources in Fig.~\ref{fig:DMsubhalo-extension}. The displayed surface brightness angular profiles have been obtained from simulated data featuring a convolution with CTAO's IRFs in the energy range from 100 GeV and 1 TeV. For definiteness, we assumed the gamma-ray spectrum of DM annihilating into $b\bar{b}$-pairs with a mass of 100 TeV for all shown sources. The selected DM sub-halos correspond to simulated objects with the median properties marked by the gray vertical lines in Fig.~\ref{fig:SL17_extensions}. In the same figure, we show the profile of a point-like source and two Gaussian-like sources with widths $0.1^{\circ}$ and $0.2^{\circ}$, respectively. These extensions are common for astrophysical objects like pulsar wind nebulae (PWNe) \cite{CTAConsortium:2023tdz}. While the angular profile of the astrophysical sources exhibits a rapid drop of brightness in agreement with their extension (and additional PSF size), the DM sub-halo profiles display a much shallower profile which continues this trend even beyond what is shown in the figure. Nonetheless, the central part of the sub-halo is about an order of magnitude brighter than its remainder. The stated angular extension of the sub-halos $\theta_{68}$ is, thus, not to be understood as a sharp boundary but rather the region from which the majority of the gamma-ray emission originates.

From these morphological considerations, we design an analysis approach to exploit spatial information of the detected subhalo to discriminate it from other source classes. The setup is as follows: We prepare CTAO mock data containing the selected background components at their nominal values predicted by the input models, i.e.~$\bm{\theta}^B = 1$, plus the single DM sub-halo at $(\ell, b) = (40^{\circ}, 0^{\circ})$. This case serves as an illustrative example for the prospects with CTAO. The DM sub-halo component is added to the mock data with variable overall normalization, which is readily translated to the DM pair-annihilation cross section (see Sec.~\ref{sec:dmsignal}).

We remain agnostic about the nature of the considered astrophysical source classes, i.e.~we do not assume realistic and, thus, model-dependent spatial morphologies linked to known TeV-bright sources like supernova remnants or PWNe. We rather consider two generic cases that we can link to conventional astrophysical sources, namely 
\begin{enumerate}
    \item a genuine point-like source (PLS) and
    \item the detection of the DM sub-halo up to a distance of 30 pc.
\end{enumerate}
The second criterion of detection up to 30 pc is motivated by the HAWC observation and detection of the two pulsar halos J0633+1746 and B0656+14 in \cite{Abeysekara:2017b} to exactly this distance. However, it does not feature an actual alternative gamma-ray emitter. The reason for including this case study is as follows: The significant detection of these two pulsar halos up to a distance of 30 pc allowed the HAWC collaboration to firmly discriminate between the pulsar halo source class and other known emitters. We expect that such distance is likewise sufficient to distinguish a DM sub-halo from conventional astrophysical sources as adopted for pulsar halos by a subset of the authors in \cite{Eckner:2022gil}. In this sense, the spatial information we obtain from such a dark subhalo is expressive enough to tell it apart from conventional sources.


This analysis is solely based on the spatial morphology of the novel source. Hence, we analyze three broad energy bands: $[0.01,0.1]$ TeV, $[0.1, 1]$ TeV, and $[1, 10]$ TeV\footnote{We refrain from adding a fourth energy band from 10 TeV to 100 TeV since it will be subdominant for all considered DM masses. Also note that integrating over the energy removes all discrimination power the dark subhalo's spectral energy distribution might add. Yet, deriving the gamma-ray spectrum already requires a high TS so that the overall gain in discrimination power is not substantial.}. To this end, we simulate the PLS candidates with the same spectrum as the DM sub-halo component. We eventually integrate the spectral information to compute the total photon counts in the three broadband energy ranges. The analysis is performed without considering the effect of systematic uncertainties. 

Lastly, we quantify the preference for the DM sub-halo model over the alternative source class in a nested model setting. The model $\bm{\mu}^{\mathrm{null}}$ of the null hypothesis consists of the selected background components as well as the conventional astrophysical source template whereas the alternative hypothesis $\bm{\mu}^{\mathrm{test}}$ adds the DM sub-halo template to it; a single renormalization parameter is assigned to each component. We derive the best-fitting parameters for both hypotheses and evaluate the TS according to Eq.~\ref{eq:tsdet}. Since the alternative hypothesis has one additional degree of freedom with respect to the null hypothesis, we find a $5\sigma$ preference for the alternative hypothesis, i.e.~the data being described by the DM sub-halo model, for $\mathrm{TS} = 25$ since the test statistic is distributed according to a $\chi^2$-distribution with one degree of freedom \cite{Cowan:2010js}.

\vspace*{11pt}
\noindent\textit{Angular decomposition of a detected DM sub-halo.} The principal idea of this analysis is to devise a sequence of gamma-ray emission models $\bm{\mu}$ that iteratively add annuli centered on the position of the detected DM sub-halo to reconstruct its gamma-ray flux up to a certain angular distance from this center. We follow the prescription of this approach in \cite{Eckner:2022gil} that was applied to pulsar halos in GPS data. The employed annuli templates are simulated from the original spatial DM sub-halo model that we divided into concentric annuli of $0.1^{\circ}$ up to 30 pc, i.e.~$1.6^{\circ}$ for our selected DM sub-halo candidate. The iterative process to determine the number of significant annuli goes as follows: We initiate the sequence of fits with $\bm{\mu}^{\mathrm{null}}$ comprised of the selected background components while the alternative model $\bm{\mu}^{\mathrm{test}}$ contains the innermost signal annulus from $0^{\circ}$ to $0.1^{\circ}$. If we obtain $\mathrm{TS} = 25$ when evaluating Eq.~\ref{eq:tsdet}, we add the annulus to the null hypothesis model $\bm{\mu}^{\mathrm{null}}$, which is employed in the next step advancing to the adjacent annulus. Otherwise, we keep the current setup of $\bm{\mu}^{\mathrm{null}}$ but enlarge the annulus of $\bm{\mu}^{\mathrm{test}}$ by adding the adjacent annulus to the one already in the model, i.e.~the annulus width is increasing; the number of fitted annuli remains the same. We continue this rationale until we reach the maximal distance from the sub-halo center of 30 pc (which is $1.6^{\circ}$ for a source at 1 kpc distance).

\subsection{Assessing the sensitivity to the entire sub-halo population}
\label{sec:details-integrated-sensitivity}

While the analysis parts outlined in the previous sections were dedicated to the CTAO prospects in an optimistic scenario, we now aim for an assessment of the CTAO's sensitivity to a dark sub-halo population within the GPS as a whole. We derive a statement about the number of detectable sources as a function of the annihilation cross section incorporating the statistical scatter of their properties and positions according to the expectations following from the 1010 realizations of the SL17 resilient and fragile populations.

W design a dedicated analysis approach that allows us to estimate the detection sensitivity to objects exhibiting a certain angular extension. The exposure of the GPS is not uniform in latitude and longitude, hence, the guiding idea is to consider the integrated energy band from 70 GeV to 100 TeV and to derive the \emph{integrated sensitivity} to a certain source type for a selected number of representative positions in the GPS band. To this end, we define a coarse tiling of the GPS ROI at 60 different locations given by 5 latitude bins with edges at $b = \left[-5^{\circ}, -3^{\circ}, -1^{\circ}, 1^{\circ}, 3^{\circ}, 5^{\circ}\right]$, and 13 longitude bins with edges at~$l = \left[-180^{\circ}, -155^{\circ}, -115^{\circ}, -85^{\circ}, -60^{\circ}, -40^{\circ}, -15^{\circ}, 15^{\circ},40^{\circ},\right.$ $\left. 60^{\circ}, 85^{\circ}, 115^{\circ}, 155^{\circ}, 180^{\circ}\right]$. The non-uniform longitude bin size is motivated by the non-uniform exposure pattern of the GPS. We avoid large overlaps between regions of strongly different exposure (c.f.~Fig.~\ref{fig:intsens}). At each position, we compute the integrated sensitivity for three source types: \emph{(i)} a point-like source, \emph{(ii)} the median representative of the SL17 fragile population and \emph{(iii)} the median representative of the SL17 resilient population. The selection of the median representative of the sub-halo populations is based on $\theta^{\ast}_{68}$ of the brightest sub-halo per realization as shown in Fig.~\ref{fig:SL17_extensions}. In that sense and to be precise, they are the median representations of the ensemble of brightest sub-halos. The gray vertical line indicates the median extension, which reads $\theta^{\ast}_{68, \mathrm{fragile}} = 0.54^{\circ}$ for SL17 fragile and $\theta^{\ast}_{68,\mathrm{resilient}} = 0.7^{\circ}$ for SL17 resilient. The other sub-halo parameters characterizing these sub-halos are stated in Tab.~\ref{tab:sub-halos}.

The integrated sensitivity represents the sensitivity to a signal integrated over a selected energy range, conveying information on the minimal integrated flux that must be emitted by the source under study for it to be detected with a $\textrm{TS} \geq 25$. The evaluation of the integrated sensitivity is done by performing the same approach as in the case of the differential sensitivity assessment, i.e.~calculating the likelihoods of the two fits (for the test and the null hypothesis) in each energy bin, but instead of iteratively assessing the normalization of the source in each energy bin, the likelihoods over all energy bins in the considered energy range are summed, and the normalization of the source is determined as the source normalization value at which the newly obtained ``integrated'' likelihood reaches $\textrm{TS} = 25$. The obtained source normalization is then used as the re-normalization factor on the initial source spectra that entered the simulations of the source template, and the integrated sensitivity, i.e., the flux that needs to be achieved for source detection, is equal to the re-normalized source spectrum integrated over the selected energy range.

Eventually, we obtain three grids of flux values that mark the threshold for the detection of our three targeted source classes at specific positions in the GPS band. Finding the required flux for claiming a significant detection regarding all three considered sources, provides us with the evolution of the CTAO sensitivity to extended sources as a function of their angular extent. Of course, the sub-halos in our population realizations are of different angular extents to what we selected as test cases. To inter- and extrapolate from these three benchmark cases, we adopt and slightly modify the approach applied in the H.E.S.S.~Galactic plane survey publication \cite{HESS:2018pbp}. In each probed tile of the GPS ROI, we parametrize the dependence of the flux sensitivity $F$ for detection as
\begin{equation}
\label{eq:flux-rescaling}
    F\!\left({\theta_{68}}\right) = A(\bm{x}) \sqrt{\alpha \theta_{68}^2 + \sigma_{\mathrm{PSF}}^2} \rm{.}
\end{equation}
The quantity $\sigma_{\mathrm{PSF}}$ refers to the energy-averaged PSF of the CTAO, which is $\sim0.05^{\circ}$ for the chosen energy band. Having derived the required flux for detecting point-like sources, we calculate the position-dependent normalization $A(\bm{x})$ for $\theta_{68} = 0$. With the flux sensitivities for $\theta^{\ast}_{68, \mathrm{fragile}}$ and $\theta^{\ast}_{68, \mathrm{resilient}}$, we optimize this re-scaling equation by varying $\alpha$ such that it minimizes the error margin at each individual position. Eventually, the minimal error we can achieve with this method is of the order of $15\%$, corresponding to a position-independent value of $\alpha = 0.06$. We derive the flux sensitivity with this re-scaling function for any sub-halo of our realizations and translate it to an annihilation cross section based on the respective $\mathcal{J}$-factor.

\section{GPS sensitivity to individual dark matter sub-halos} \label{sec:sensitivity_single}

In this section, we discuss the obtained results on the sensitivity of the CTAO's GPS to the most optimistic scenario with the brightest sub-halos found across our simulations being located in the Galactic plane as described in Sec.~\ref{sec:single_SH}. An exploration of the prospects for the full population follows in Sec.~\ref{sec:population}.

\subsection{Sensitivity to sub-halos from a spectral analysis}\label{sec:flux-sensitivity}

\begin{table*}
    \begin{tabular}{p{3cm} p{2.cm} p{2.cm} p{5.cm} p{2.cm} p{2.cm}}
    \hline \hline
    Sub-halo mass $M_{\Delta}$ & Distance & $r_s$ & $\mathcal{J}_{\rm tot}$ [GeV$^2$ cm$^{-5}$] & $r_{\Delta}$ & $\theta_{68}$ \\  [0.5ex]
    \hline
    $4.8\cdot10^6 M_{\odot}$ & 0.92 kpc & 2.36 kpc & 9.56 $\cdot 10^{21}$ (6.44 $\cdot 10^{21}$ in FoV) & 0.162 kpc & 1.54$^\circ$ \\
    $4.8\cdot10^6 M_{\odot}$ & 5.0 kpc & 2.36 kpc & $3.21 \cdot 10^{20}$ & 0.162 kpc & 0.46$^\circ$ \\
    $4.8\cdot10^6 M_{\odot}$ & 10.0 kpc & 2.36 kpc & $8.02 \cdot 10^{19}$ & 0.162 kpc & 0.23$^\circ$\\
    $4.8\cdot10^6 M_{\odot}$ & 30.0 kpc & 2.36 kpc & $8.94 \cdot 10^{18}$ & 0.162 kpc & 0.08$^\circ$\\
    \hline
    $8.9\cdot10^5 M_{\odot}$ & 1.12 kpc & 1.18 kpc & 9.72 $\cdot 10^{20}$ & 0.099 kpc & 1.24$^\circ$\\
    $7.9\cdot10^4 M_{\odot}$ & 0.95 kpc & 0.29 kpc & 1.71 $\cdot 10^{20}$ & 0.040 kpc & 0.57$^\circ$\\
    \hline
    $7.7\cdot10^7 M_{\odot}$ & 12.3 kpc & 1.26 kpc & $8.94 \cdot 10^{19}$ & 1.02 kpc & 0.7$^\circ$\\
    $1.1 \cdot 10^7 M_{\odot}$ & 11.6 kpc & 0.529 kpc & $1.59 \cdot 10^{19}$ & 0.54 kpc & 0.54$^\circ$\\
    \hline
    \end{tabular}
    \caption{Table of selected sub-halos simulated with \texttt{CLUMPY} and considered in the assessment of the GPS sensitivity, listing in consecutive columns the final (after tidal stripping) sub-halo mass $M_{\Delta}$, its distance from Earth, its scale radius $r_s$, the total $\mathcal{J}_{\rm tot}$-factor values, representing the $\mathcal{J}$-factor of the sub-halo across its full extent (up to $r_{\Delta}$), and the $\mathcal{J}_{\rm FoV}$-factor values, representing the fraction of the $\mathcal{J}_{\rm tot}$-factor observed within the telescope's FoV, the total radial size $r_{\Delta}$, and the angular size $\theta_{68}$ of the sub-halo. The first four rows list different instances of the overall brightest sub-halo found across all realizations and population scenarios at increasing distances from Earth. The first entry corresponds to the original dark sub-halo from the SL17 resilient scenario. The next three rows list sub-halo realizations found in the SL17 resilient populations around a distance of 1 kpc selected to quantify the impact of decreasing $r_{\Delta}$. The last two entries list the parameters of the median SL17 resilient and fragile representatives marked by grey lines in Fig.~\ref{fig:SL17_extensions}.  \label{tab:sub-halos}}
\end{table*}

\subsubsection{Flux sensitivity for GPS conditions}\label{sens:flux}

\begin{figure*}[t]
\begin{centering}
\includegraphics[width=0.60\linewidth]{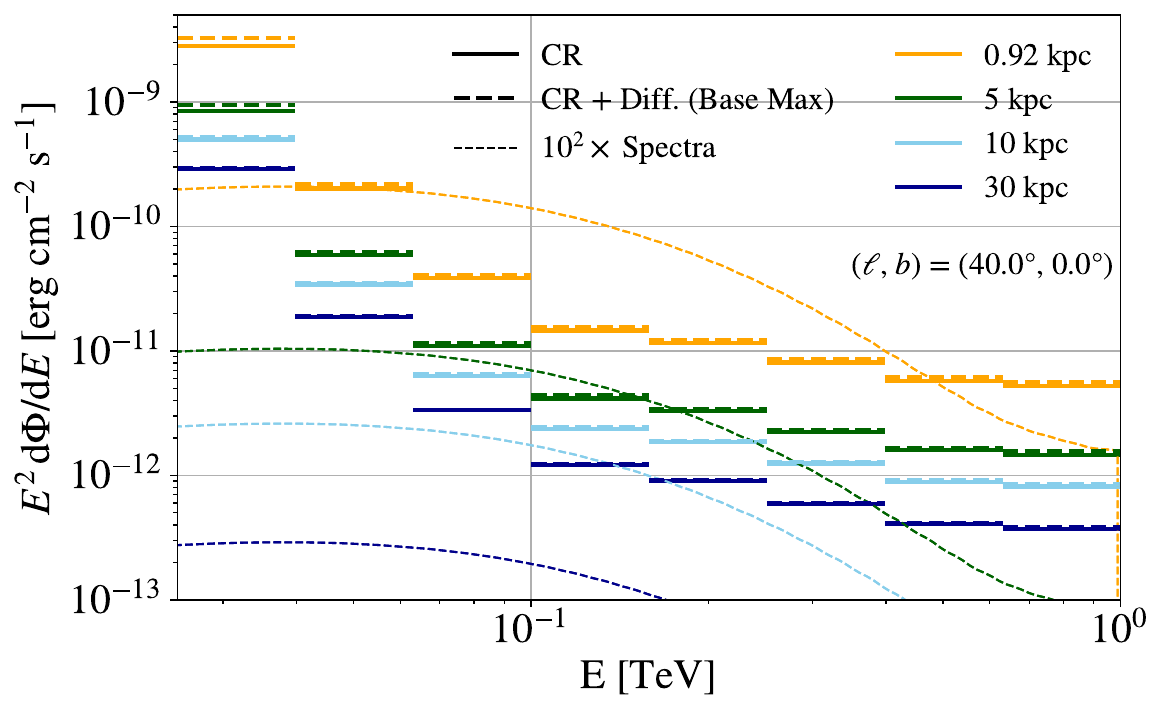}
\par\end{centering}
\caption{The differential sensitivity, defined as the minimum flux needed to obtain a 5$\sigma$ detection of the benchmark sub-halos positioned at four different distances from the observer and at the reference positions $(l, b) = (40.0^\circ, 0.0^\circ)$. 
The sensitivity does not simply scale with the inverse of the distance squared due to substantial angular extensions for sub-halos that exceed the latitude extension of the GPS, which causes only a fraction of the total flux to be contained within the ROI. The sensitivities are calculated with (dashed lines) and without the IE (solid lines). The thermal DM spectra for each sub-halo are overlaid and enhanced by two orders of magnitude to fit within the plotted flux range (following the same color code as the sensitivity lines). The $\mathcal{J}$-factor values for each sub-halo position and distance are listed in Table \ref{tab:sub-halos}.}
\label{fig:fluxsens}
\end{figure*}

The spectral sensitivities, shown in Fig. \ref{fig:fluxsens}, correspond to the benchmark brightest sub-halo in the simulated populations of sub-halos positioned at different distances from Earth (the first section of the sub-halo list shown in Tab.~\ref{tab:sub-halos}). The sensitivity degrades with decreasing distance to the sub-halo which corresponds to a larger source extension (see the last column in Tab. \ref{tab:sub-halos} for the angular size of sub-halos positioned at various distances). 



Fig.~\ref{fig:fluxsens} displays the flux levels of our reference single sub-halo model (dotted lines) scaled to higher values by two orders of magnitude to allow the flux to fall within the plotted flux range for the obtained sensitivities. The fluxes are computed assuming the thermal cross section for DM annihilation, $\langle\sigma v\rangle = 3\cdot 10^{-26}$ cm$^3$ s$^{-1}$. We note while this numerical value for the thermal cross section is commonly taken as a reference, there are various particle physics models of DM in which the current annihilation rate can be larger than in the early universe. Such scenarios may occur in the presence of
resonances \cite{Griest:1990kh, Baer:2003bp, Kakizaki:2005en, Ibe:2008ye, Arina:2014fna} or they may be related to the so-called Sommerfeld effect \cite{Hisano:2004ds}. The expected annihilation cross section can reach levels more than one order of magnitude above the vanilla WIMP value.  

\begin{figure*}[t]
\begin{centering}
\includegraphics[width=0.48\linewidth]{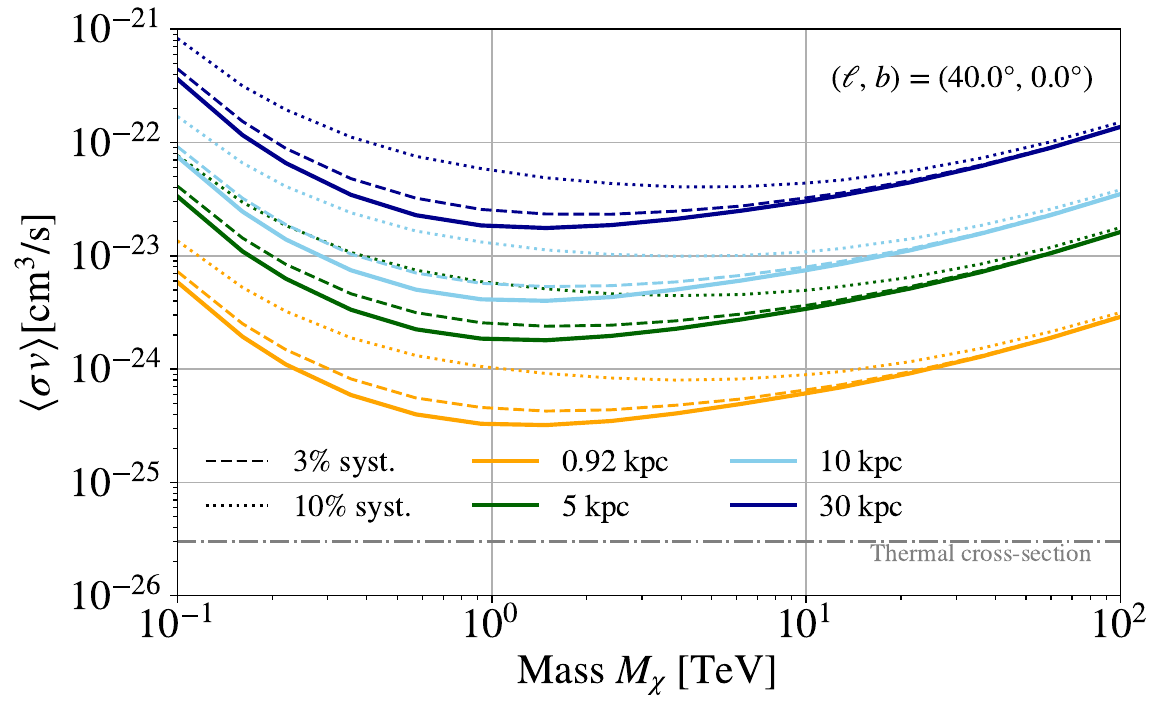}\hfill
\includegraphics[width=0.48\linewidth]{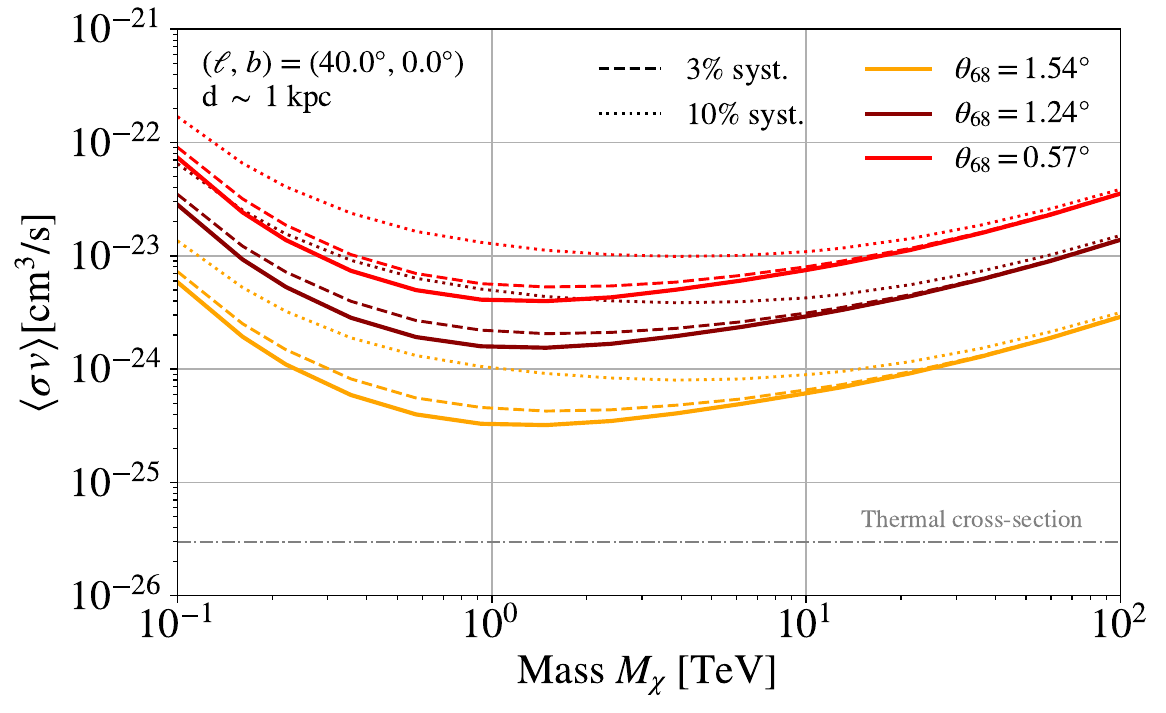}
\par\end{centering}
\caption{Sensitivity of the CTAO GPS to the brightest DM sub-halo candidate found among all realizations, described in Sect. \ref{sec:model} and listed in Tab. \ref{tab:sub-halos}, with respect to the DM particle mass $M_\chi$. The benchmark sub-halos are positioned at the reference location within the planned GPS at $(\ell, b) = (40.0^\circ, 0.0^\circ)$. (\emph{Left}:) Sensitivity to the brightest sub-halo positioned at four distances from Earth, assuming no systematic uncertainty (solid lines) or various levels of added systematics (dashed, and dotted lines). \emph{Right:} Sensitivity to the brightest sub-halos with different extensions, positioned at the reference distance of 1 kpc as found in the sub-halo population realizations.
}
\label{fig:sigmavM}
\end{figure*}

To explore the cross sections that would be within the reach of the GPS, we analyze the spectra of annihilating DM with masses ranging from $0.1 - 100$ TeV. The results for the brightest sub-halo positioned at various distances, shown in the left panel of Fig.~\ref{fig:sigmavM}, suggest that GPS will achieve the best sensitivity to DM masses in the $0.5 - 3$ TeV range, and for sources positioned closest to the observer, reaching the $\langle\sigma v\rangle$ values down to $\sim 3\times10^{-25}$ cm$^3$ s$^{-1}$. This is, however, only true in case no systematic uncertainty is considered in the analysis. The added systematic uncertainty significantly modifies the CTAO reach in the range where it is most sensitive, leading to significantly different results (see Sec.~\ref{sec:sens_iemsyst} for further discussion). Another important thing to note is that the flux sensitivity estimate is also annihilation channel-dependent. Here, as mentioned before, we only consider the annihilation into $b\bar{b}$ final states. 

We emphasize the striking fact that the ordering of lines in Fig.~\ref{fig:fluxsens} and the left panel of Fig.~\ref{fig:sigmavM} is inverted. The brightest sub-halo at $\sim 1$ kpc distance requires the lowest annihilation cross section for detection while it requires the highest differential flux among all probed sub-halos for a detection. The reason for this observation is the $\mathcal{J}$-factor, which compensates for the loss in differential sensitivity caused by the large angular extension of the sub-halo. In general, the ordering of sensitivity in Fig.~\ref{fig:fluxsens} is driven by the angular size of the sub-halo since it is easier to detect a point-like source (far distant sub-halo) than an extended (close) one. The $\mathcal{J}$-factor is suppressed by a factor of $1/d^2$ so that sub-halos with large $\mathcal{J}$-factor are typically closer to the observer.

The reach of the GPS in the case of sub-halo models with varying $r_{\Delta}$/$\theta_{68}$ listed in the central section of the sub-halo properties Tab.~\ref{tab:sub-halos} is shown in the right panel of Fig. \ref{fig:sigmavM}. It also features the overall brightest sub-halo (same line color as in the left panel). They represent the sub-halos positioned at the distance of $\sim 1$ kpc selected from the simulated sub-halo population realizations. For sub-halos with different extensions but roughly equal distances, the most constraining limits are found for the overall brightest sub-halo since, in this case, the total $\mathcal{J}$-factor is directly linked with the physical extension. One can interpret this scenario as a bracketing study for the impact of varying the properties of the brightest sub-halo in the population. The properties of nearby sub-halos may vary to such an extent that the required detection flux deteriorates by around one order of magnitude. Therefore, the yellow lines in Fig.~\ref{fig:sigmavM} are indeed the most optimistic expectations regarding the sensitivity of CTAO.

From the presented results, and as we will see in Sec.~\ref{sec:population}, the reach of GPS allows for the detection of sub-halos when cross sections are at best one order of magnitude above the thermal cross section.

\begin{figure*}[t]
\begin{centering}
\includegraphics[width=0.50\linewidth]{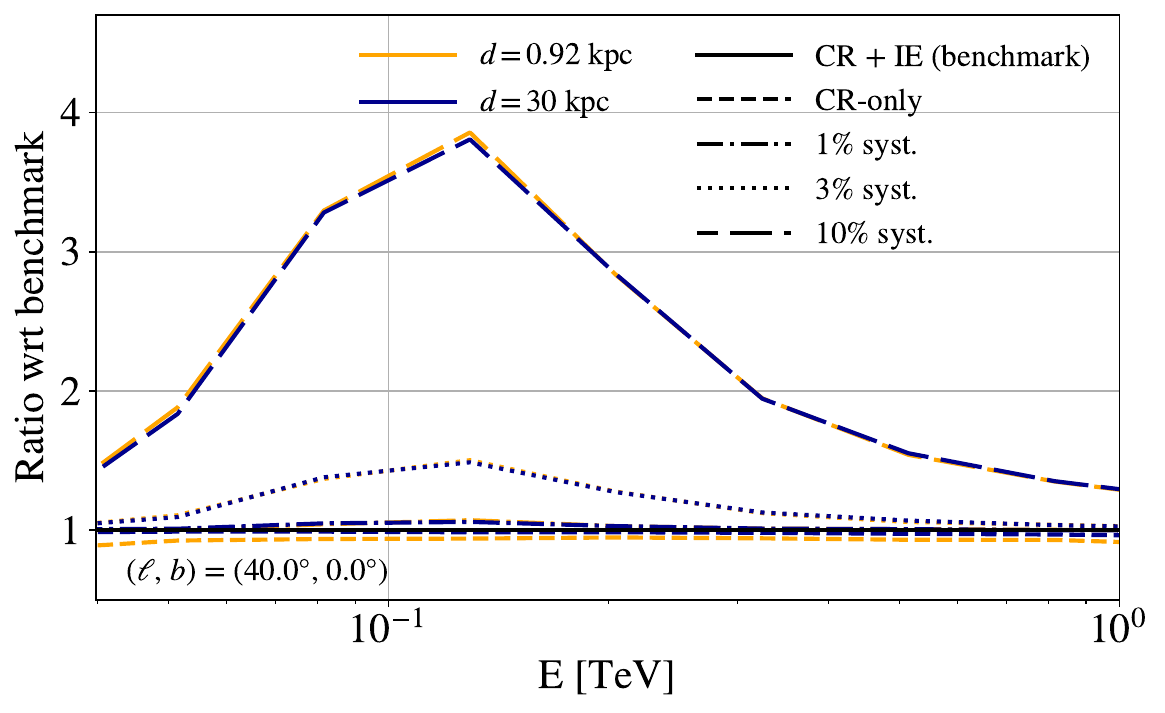}
\par\end{centering}
\caption{The differential sensitivity ratio with respect to the benchmark model, showing the effect of the IE (dashed lines) and two different levels of systematic uncertainty (1\%, 3\% and 10\% levels are shown with dash-dotted, dotted and wide-dashed lines, respectively) on derived sensitivities to sub-halos positioned at 920 pc and 30 kpc distance in yellow and blue colors, respectively. The benchmark model takes into consideration the instrumental (CR) and the IE, without added systematics.
}
\label{fig:fluxsens_ratio}
\end{figure*}


\subsubsection{Impact of IE and systematic uncertainties}\label{sec:sens_iemsyst}

From the results presented in Figs.~\ref{fig:fluxsens} and \ref{fig:sigmavM}, we can assess the impact of the IE and systematics, respectively. While including the IE in our analysis does not significantly affect the derived sensitivities, the same cannot be said about the effect of systematics, which is most significant precisely at the energies where CTAO would probe the lowest $\langle\sigma v\rangle$ values. The degree to which the systematic uncertainty worsens the derived sensitivity strongly depends on the overall systematic error amplitude included in the model. We explore three different levels of systematic uncertainty: 1\%, 3\%, and 10\%. While the 1\% added systematics do not significantly modify the conclusions that we can emanate from the results that do not consider systematic uncertainty, the effects become noticeable at the 3\% level, and cause the obtained sensitivities at energies below few TeV to be dominated by systematics at the 10\% level. There is no benchmark model for systematic uncertainty for CTAO at this time, however, we consider in our selection of the explored systematic errors that the target level of systematic uncertainty for CTAO is expected to be at the level of a few per cent.

As another way to look at these results and to facilitate the comparison of different levels of systematic uncertainty, we display their relative impact in Fig.\ref{fig:fluxsens_ratio} for two cases of DM sub-halo templates from Fig.~\ref{fig:fluxsens}; located at a distance of either $d=0.92$ kpc or $d = 30$ kpc. The impact is shown relative to the differential flux sensitivity without systematic uncertainties and a background comprised of CR and IE. We see that the inclusion of IE has the strongest impact on a nearby source since it will appear as a very extended object that can be affected by the structure of the IE. A distant sub-halo rather classifies as a PLS stays nearly unaffected by this additional background component. It becomes likewise evident that systematic uncertainties at the level of 3\% already deteriorate the expected sensitivity by more than 40\% for energies around 100 GeV. Note that this energy range dominates the sensitivity to DM particles of masses around 1 TeV.

\subsection{Discrimination from astrophysical sources and intensity profile reconstruction}
\label{sec:source-discrimination}

We aim to assess the following scientific question: Given the detection of a novel localized gamma-ray emitter above the expected CTAO-relevant backgrounds in GPS data how can we determine the nature of the new source? Of course, our objective is targeting DM sub-halos and CTAO's capabilities of discriminating between this exotic class of gamma-ray sources and conventional emitters \emph{using only the gamma-ray data}. 
Naturally, for some of the sources discovered by the CTAO, the gamma-ray source will be spatially coincident with sources detected in other electromagnetic wavelengths and therefore unlikely to be of a DM origin.

\vspace*{11pt}
\noindent\textit{Morphological discrimination.} To answer this question, we employ the rationale outlined in Sec.~\ref{sec:single-halo-analysis-details}. In particular, we prepare CTAO mock data containing the irreducible CR background and our fiducial model of IE at their nominal values predicted by the input models plus the single DM sub-halo representing the brightest candidate among the populations found in the simulations. To this end, we use the sub-halo listed in the top row of Tab.~\ref{tab:sub-halos} placed at about 1 kpc distance from Earth that corresponds to the brightest sub-halo among all simulated realizations. This case serves as an illustrative example for the prospects with CTAO; changing scale radius, sky position, or distance to the observer will have an impact analogous to the findings obtained with the spectral sensitivity analysis in Fig.~\ref{fig:fluxsens}. We refer the reader to Sec.~\ref{sec:flux-sensitivity} for a more explicit discussion of the impact of these parameters. The DM sub-halo component is added to the mock data with adjustable thermally averaged annihilation cross section of DM pair-annihilation into $b\bar{b}$ final states for DM masses ranging from 100 GeV to 100 TeV.

\begin{figure*}[t]
\begin{centering}
\includegraphics[width=0.49\linewidth]{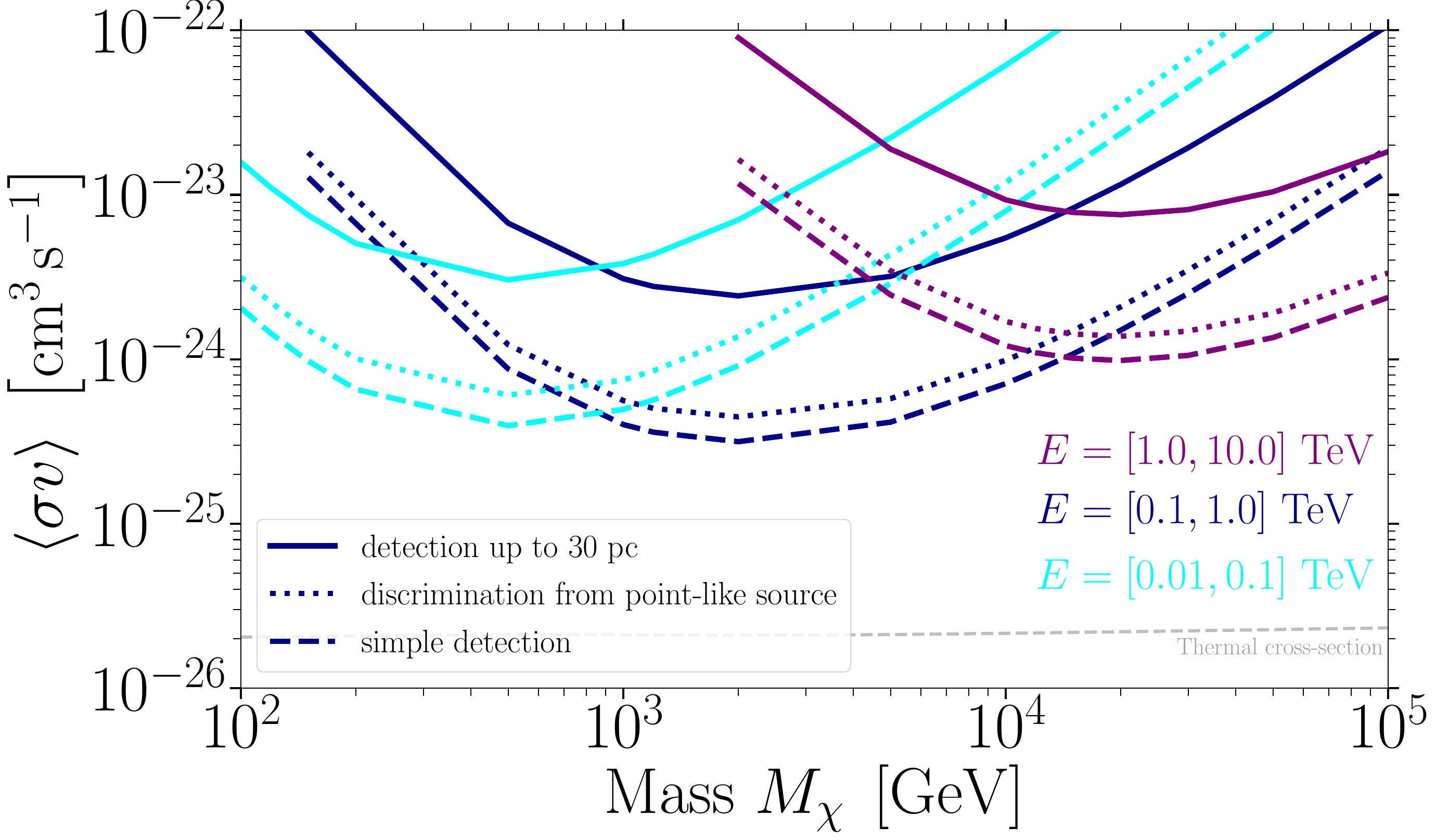}\hfill
\includegraphics[width=0.465\linewidth]{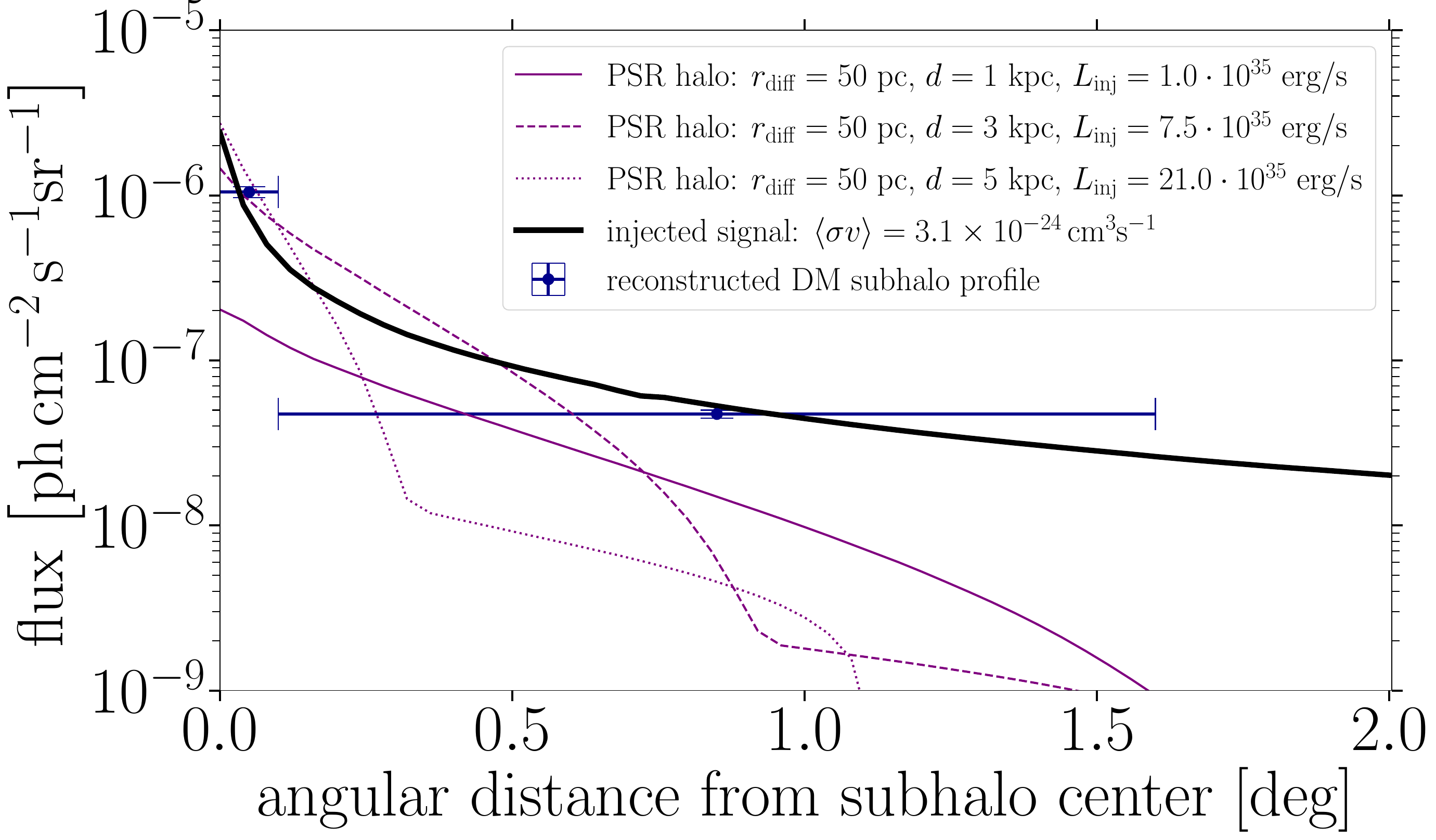}
\par\end{centering}
\caption{(\emph{Left}:) Minimal thermally averaged annihilation cross section $\langle\sigma v\rangle$ required to discriminate between a point-like source and a DM sub-halo as a function of DM particle mass for three energy bands: $[0.01, 0.1]$ TeV (cyan), $[0.1, 1]$ TeV (blue), and $[1, 10]$ TeV (purple). The fiducial DM sub-halo is located at $(\ell, b) = (40.0^{\circ}, 0.0^{\circ})$, about 1 kpc away. Results are based on CTAO mock data, including CR background and our benchmark IE. Dashed lines show the cross section for sub-halo detection, and dotted lines for discriminating the sub-halo from a point source. The solid line indicates the cross section required to resolve the DM sub-halo into at least two significant annuli up to 30 pc from its center. The dashed gray line marks the cross section for thermal DM production. (\emph{Right}:) Angular decomposition of the DM sub-halo in the 100 GeV to 1 TeV band, assuming a DM mass of 1 TeV and annihilation into $b\bar{b}$ with $\langle\sigma v\rangle = 3.1 \cdot 10^{-24}$ cm$^3$s$^{-1}$. The blue data points show the decomposition using annuli of $0.1^{\circ}$, with the input DM sub-halo profile shown in black. Error bars indicate statistical uncertainty. The DM profile is compared to ``Geminga-like'' pulsar halos at different distances, with a diffusion suppression radius $r_{\mathrm{diff}} = 50$ pc \protect\cite{Eckner:2022gil}. The pulsar halos are normalized to match the DM sub-halo flux in the 100 GeV to 1 TeV band.
\label{fig:angular_decomposition}}

\end{figure*}

We recall from the description of our analysis approach that we consider two generic cases that we can link to conventional astrophysical sources, namely 
\begin{enumerate}
    \item a PLS and
    \item the detection of the DM sub-halo up to a distance of 30 pc (or $\sim 1.6^{\circ}$ for this object at 1 kpc distance).
\end{enumerate}
For each DM mass and energy band, we scan the range of $\langle\sigma v\rangle$ to determine the value that eventually yields this threshold value of the TS required for significant discrimination. The results of this scan are reported in the left panel of Fig.~\ref{fig:angular_decomposition}, in which we also show as dashed lines the cross section required to claim a simple detection of a gamma-ray source (null hypothesis: CR background and IE, alternative hypothesis: same as the null hypothesis plus DM sub-halo).

\vspace*{11pt}
\noindent\textit{Detection up to 30 pc.} We combine this analysis part with the question of when can we expect to extract the spatial morphology of the detected source. To this end, we employ the strategy for the decomposition of a detected DM sub-halo outlined in Sec.~\ref{sec:single-halo-analysis-details}. We apply this rationale until we reach the maximal angular distance from the sub-halo center of $1.6^{\circ}$ for the chosen subhalo at around 1 kpc distance, which corresponds to a physical distance from the sub-halo's center of 30 pc. Depending on the chosen annihilation cross section of the DM sub-halo in the mock data, we obtain a certain number of significant annuli that can be used to infer the parameters of the underlying DM density profile. For us, at least two significant annuli are necessary to extract information about the DM density profile. In the left panel of Fig.~\ref{fig:angular_decomposition}, we report as solid lines the minimal annihilation cross section values that allow for decomposing the sub-halo in at least two such annuli.

\vspace*{11pt}
\textit{Results.} Focusing first on the case of simple DM sub-halo detection, we find annihilation cross section values comparable to the orange line shown in the left panel of Fig.\ref{fig:fluxsens}. We do not expect a one-to-one correspondence in the first place due to the differences in the analysis approach. The threshold cross section shown in the previous section require a TS value of 25 in any of the energy bins within the considered full range from 10 GeV to 10 TeV while here we show the prospects for integrated emission. In general, CTAO's capability to detect DM sub-halos appears to be driven by the energy range from 10 GeV to 1 TeV. Discriminating the DM sub-halo from a PLS necessitates cross section about a factor of 1.5 larger than the ones allowing for simple detection. 

In contrast, it turns out that a meaningful decomposition and detection up to 30 pc requires an annihilation cross section whose value is at least one order of magnitude higher than that required to detect the sub-halo in the first place. For a DM mass around the TeV scale, this implies cross section values  $\sim3\times10^{-24}$ cm$^{3}$ s$^{-1}$. An example of the case of DM pair-annihilation into $b\bar{b}$ final states and a mass of $M_{\chi} = 1$ TeV with a cross section at this threshold value is shown in the right panel of Fig.~\ref{fig:angular_decomposition}. The theoretical DM density profile is displayed as a solid black line while the reconstructed annuli are depicted as blue data points; the vertical error bars quantify the statistical uncertainty of the fit. A common feature in the setting of only two reconstructed annuli is the fact that the innermost annulus up to $0.1^{\circ}$ from the center is always significantly detected whereas the second annulus is extremely extended. This observation is a reflection of the rather steep density profile of the NFW profile, which renders it rather distinct from conventional astrophysical sources. Once CTAO observational data is available and an extended source has been firmly detected, the underlying spatial profile of the emission can already be assessed to some extent. Thus, the experimental approach to derive the best-fitting spatial profile requires a scan of different DM density profile parameters and subsequent angular decomposition of the detected extended signal with respect to these profiles. However, the involved and intricate fitting process needed in practice is beyond the scope of this work.

In the age of CTAO, the search for pulsar halos will make significant progress. Few examples of this new class of gamma-ray emitters are known to date \cite{Liu:2022hqf}. Since we have seen in the previous section that the emission profile of a DM sub-halo is different from conventional (generic) spatial profiles, it is interesting to ask whether a newly discovered source is a pulsar halo or, even more excitingly, produced by DM annihilation. In Fig.~\ref{fig:angular_decomposition} we compare the fiducial case of gamma-ray emission from a DM sub-halo with different instances of pulsar halos (purple lines). The models are adopted from \cite{Eckner:2022gil} where we select the set of Geminga-like halos and we fix the size of the zone of reduced diffusion efficiency to $r_{\mathrm{diff}} = 50$ pc; the benchmark case in the referenced publication. The different line styles denote pulsar halos at increasing distances from Earth. The luminosities of the pulsar halo examples are chosen to reproduce the same integrated flux from 100 GeV to 1 TeV as the DM sub-halo model. Note that the pulsar halo at $d = 1$ kpc undershoots the DM sub-halo flux profile because the spectrum of both models is different. We have shown in \cite{Eckner:2022gil} that the stated injection powers $L_{\mathrm{inj}}$ for all three pulsar halo examples are high enough to guarantee a meaningful angular decomposition of the respective source up to at least 30 pc from their centers with more than two significant annuli (see also Fig.~5 in \cite{Eckner:2022gil}). The chosen DM sub-halo template already corresponds to the most optimistic scenario and given the results in the left panel of Fig.~\ref{fig:angular_decomposition} also the chosen DM mass provides the best prospects for an angular decomposition\footnote{The required cross section for detection up to 30 pc is around the minimum ($\sim 3\cdot10^{-24}$ cm$^3$ s$^{-1}$ at $M_{\chi} = 2$ TeV) of the profile.}. Therefore, any other dark sub-halo will most likely require larger cross sections, i.e.~a higher luminosity, for a meaningful decomposition into at least two annuli. In return, pulsar halos with the same overall luminosity may be even more finely decomposed. Thus, we conjecture that for an associated detected source characterized by a luminosity such that a putative dark sub-halo could be meaningfully decomposed into annuli, it is feasible to tell apart the possibility of it being a pulsar halo or a genuine sub-halo.


\section{GPS sensitivity to a population of dark matter sub-halos} \label{sec:population}

\subsection{Accessible fraction of the sub-halo population}\label{sec:pop-study}

\begin{figure}[t]
\begin{centering}
\includegraphics[width=0.9\linewidth]{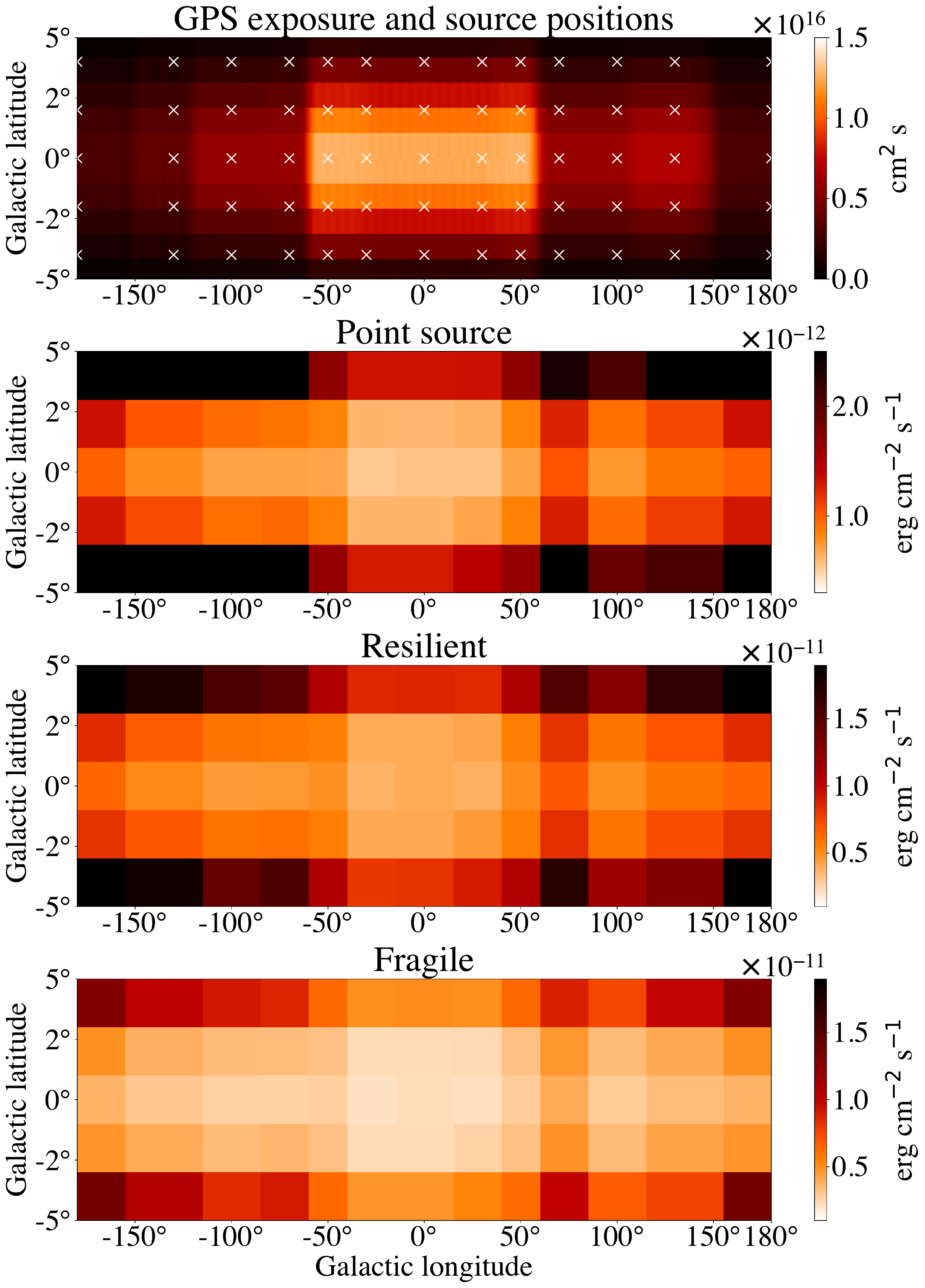}
\par\end{centering}
\caption{The integrated sensitivity to sub-halo models with various extensions, integrated over the energy range $0.07-100$ TeV. The positions of the sub-halos are shown in the upper panel with white marks, and their distribution is overlaid with the exposure map of the CTAO Galactic plane survey region. The lower three panels show the integrated sensitivity to a point source, a sub-halo with a mean extension of the sub-halos in SL17 resilient populations, and a sub-halo with a mean extension of the sub-halos in SL17 fragile populations, in this order.}
\label{fig:intsens}
\end{figure}

As outlined in Sec.~\ref{sec:details-integrated-sensitivity}, we analyze the accessible fraction of a sub-halo population via position-dependent integrated sensitivities for the GPS band. These sensitivity maps are obtained for fragile, resilient, and point source templates and displayed in Fig.~\ref{fig:intsens}.  To identify whether a sub-halo is detectable or not, we compute the total integrated flux of the sub-halo using Eq.~\ref{eq:integr-flux}. Based on the location of the sub-halo in the GPS region as found in our population realizations, its total integrated flux is compared with the expected CTAO sensitivity at that location from applying the re-scaling formula in Eq.~\ref{eq:flux-rescaling} taking into account its angular extent $\theta_{68}$. We interpolate to any intermediate position of the entire GPS region using the cubic-kind 2d-interpolation. If the total integrated flux is greater than the derived sensitivity at this position then the sub-halo is classified as detected.

\begin{figure*}[t!]
\centering
\includegraphics[width=\columnwidth]{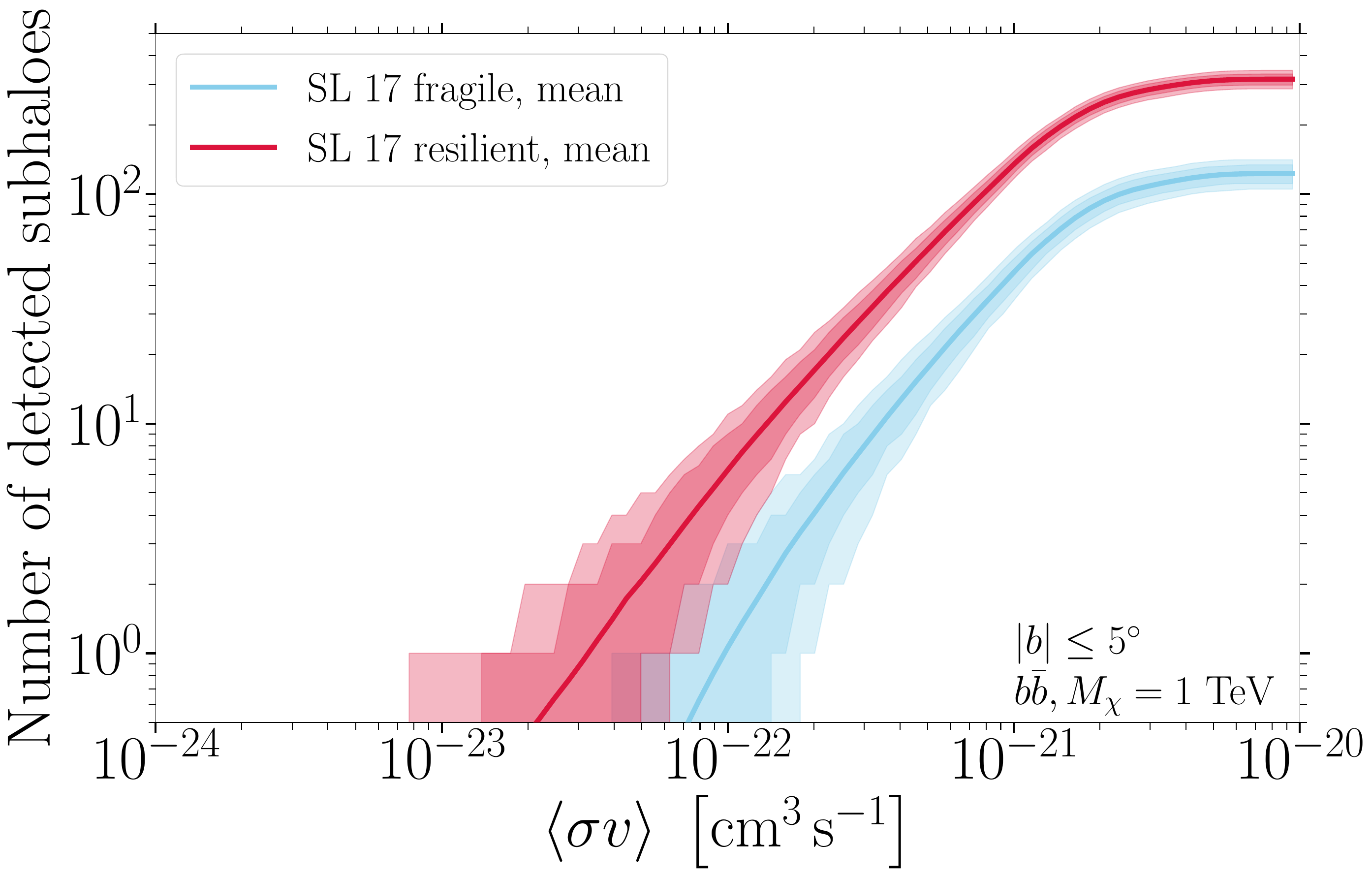}\hfill
\includegraphics[width=\columnwidth]{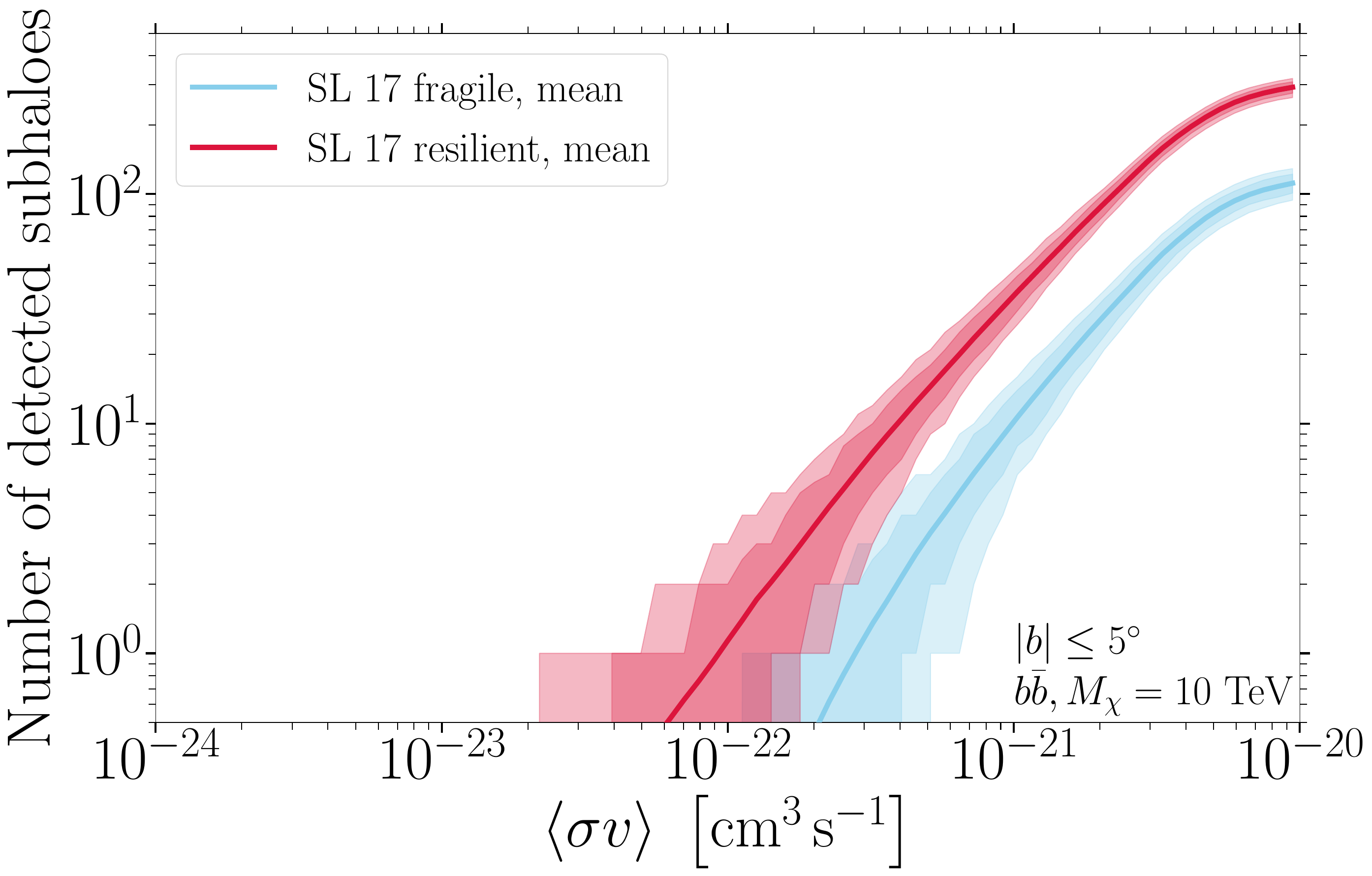}
\caption{Number of detectable DM sub-halos for the two population models SL17 resilient (red) and SL17 fragile (blue). The sensitivity derives from the integrated sensitivity in Fig.~\ref{fig:intsens} re-scaled via Eq.~\ref{eq:flux-rescaling} to the angular extent of each probed sub-halo accounting for the position-dependent threshold flux of the GPS assuming a DM particle of mass $M_{\chi} = 1$ TeV (\emph{left}) or $M_{\chi} = 10$ TeV (\emph{right}) annihilating into $b\bar{b}$ final states. The mean numbers are derived from the 1010 realizations for each population scenario selecting only sub-halos at Galactic latitudes $|b|\leq5^{\circ}$. The colored shaded bands indicate the 1$\sigma$ and $2\sigma$ containment bands reflecting the scatter of the numerical simulations. The mean cross section to detect at least one subhalo is $\langle\sigma v\rangle = 3.3^{+1.7}_{-1.9}\,(9.3^{+4.4}_{-4.8}) \times10^{-23}$ cm$^3$/s for SL17 resilient and $\langle\sigma v\rangle = 9.7_{-5.4}^{+4.9}\,(28^{+14}_{-13}) \times10^{-23}$ cm$^3$/s for SL17 fragile at $M_{\chi} = 1 (10)$ TeV. The stated numerical uncertainties refer to the $1\sigma$ containment band.}
\label{fig:SL17_population_detection}
\end{figure*}

We apply this formula to all sub-halos in the 1010 realizations of SL17 resilient and fragile, respectively. However, we limit the considered set of sub-halos to those falling into the Galactic plane with $|b|\leq5^{\circ}$. As a function of annihilation cross section, our approach yields a quantitative estimate of how many sub-halo of each population can be detected. The results are visualized in Fig.~\ref{fig:SL17_population_detection}. The solid lines indicate the mean evolution of the number of detected objects for SL17 resilient (red) and SL17 fragile (blue). We also provide the scatter of these quantities displayed as colored shaded bands where the high opacity band denotes the 68\% containment and the lighter band the 95\% containment.

As a general trend, we confirm that the SL17 resilient population produces more sub-halos within the GPS ROI due to their resilience to tidal shocks and stripping. To detect at least one sub-halo of this population, the annihilation cross section for a WIMP particle of mass $M_{\chi} = 1$ TeV needs to amount to $\langle\sigma v\rangle = 3.3\times10^{-23}$ cm$^3$/s on average. The SL17 fragile population is fainter on average requiring a cross section value of $\langle\sigma v\rangle = 9.7\times10^{-23}$ cm$^3$/s to allow for a single detected sub-halo. However, the scatter of $\mathcal{J}$-factors for the resilient population is more pronounced than for the fragile one. There are realizations in SL17 resilient simulations where a cross section close to $10^{-24}$ cm$^3$/s is already sufficient for detection. Thus, we are closer to the findings in Sec.~\ref{sec:sensitivity_single} that examine the most optimistic scenario for CTAO.

\subsection{Imprint of the unresolved sub-halo population}\label{sec:popstudy-subthresh}



A DM sub-halo population in the MW generates a certain level of gamma-ray emission irrespective of the potential detectability of individual sub-halos. Their cumulative emission is part of the diffuse Galactic emission whose main components are IE and the part of gamma-ray source class populations too dim to be resolved individually. In fact, DM sub-halos may be considered an exotic contribution to the latter component. To estimate the importance of this exotic diffuse contribution we compare the cumulative flux of our two DM sub-halo population scenarios with expectations for TeV-bright astrophysical sources in the GPS. 

\begin{figure*}[t]
\begin{centering}
\includegraphics[width=0.49\linewidth]{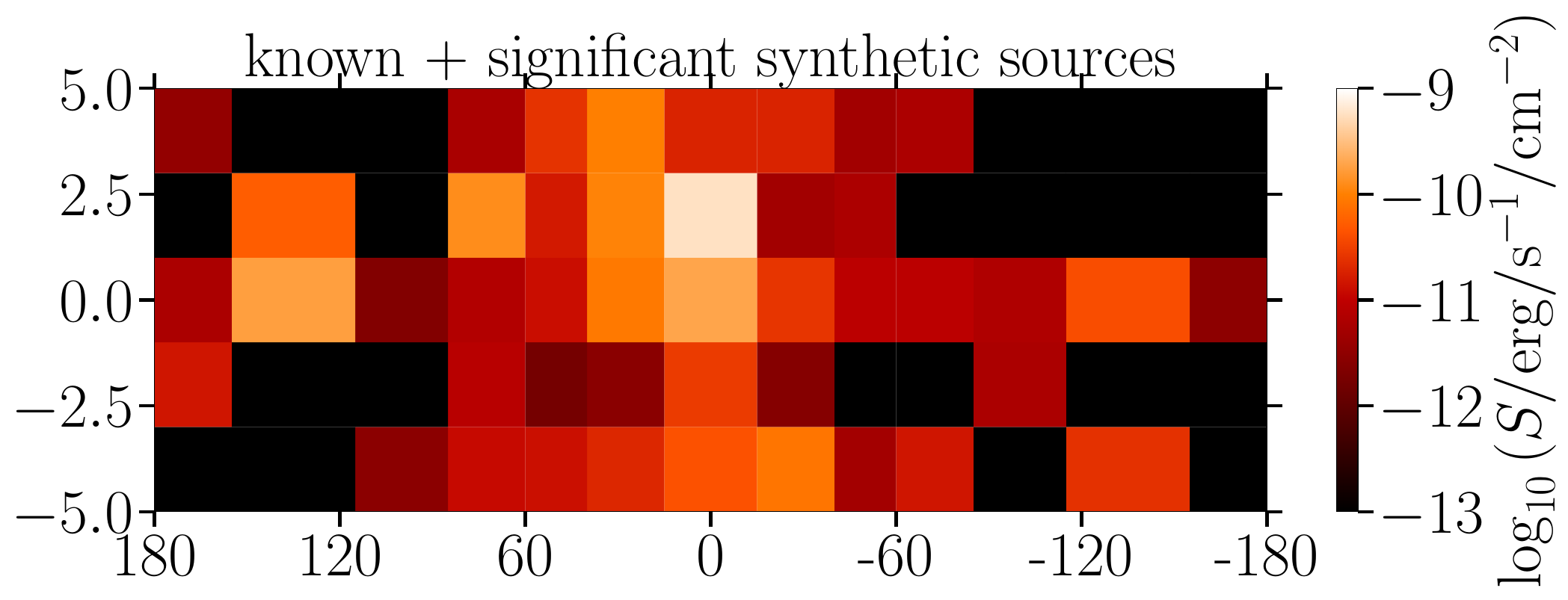}\hfill
\includegraphics[width=0.49\linewidth]{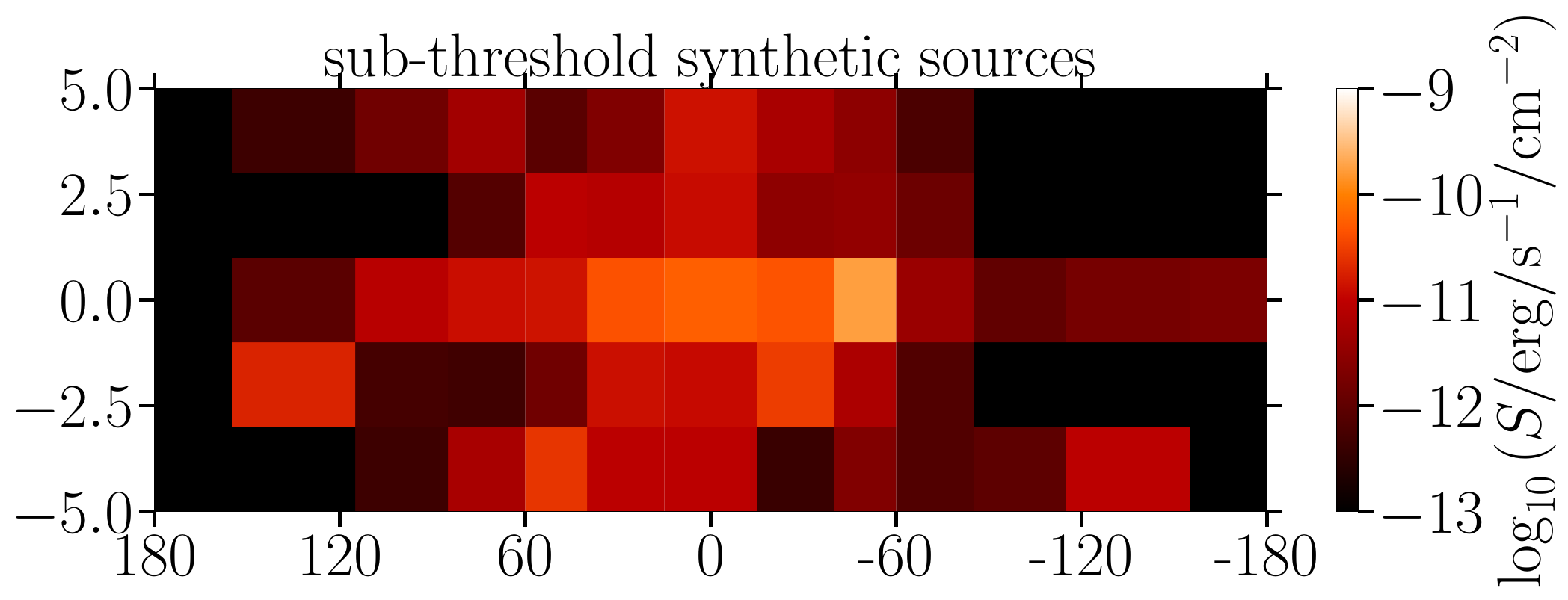}\\
\includegraphics[width=0.49\linewidth]{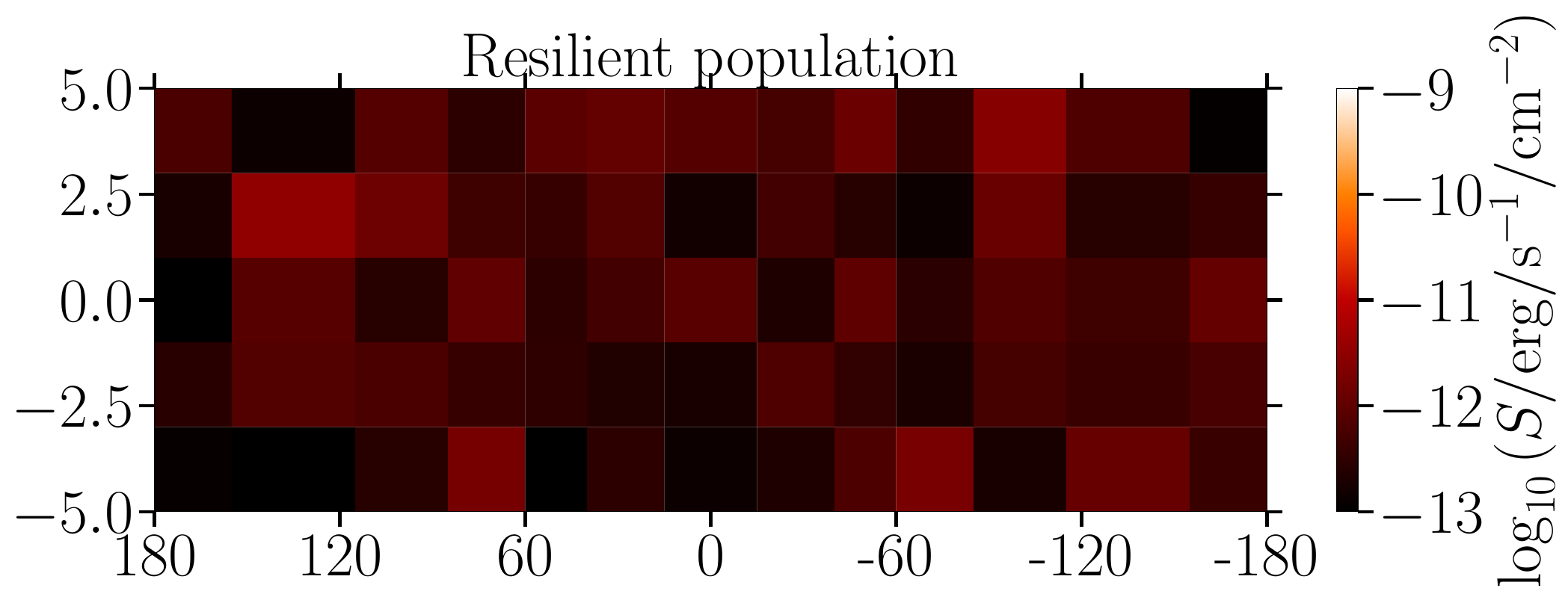}\hfill
\includegraphics[width=0.49\linewidth]{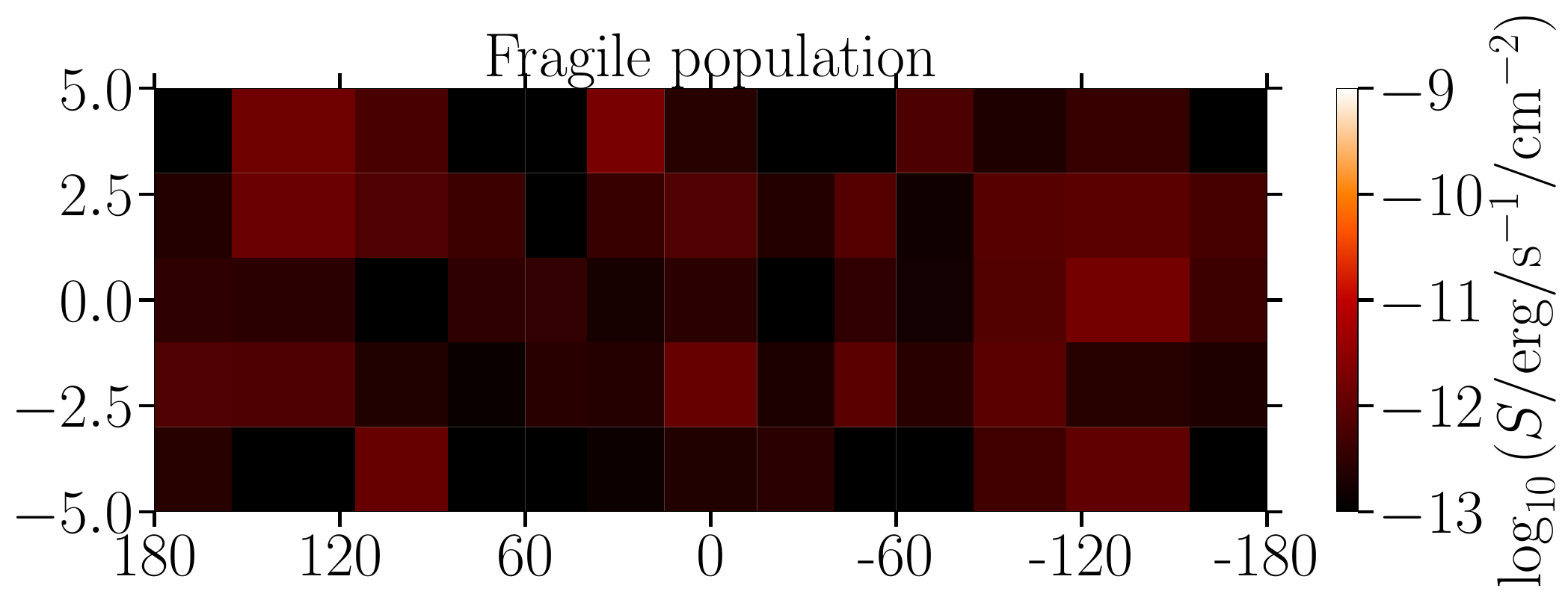}

\par\end{centering}
\caption{(\emph{Top row}:) Cumulative intensity $S$ of the source population comprised of known and synthetic PWNe sources simulated and employed in CTAO's GPS publication \cite{CTAConsortium:2023tdz}. The intensity map follows the spatial binning of Fig.~\ref{fig:intsens} and has been calculated per source from 70 GeV to 100 TeV. The left panel displays the cumulative emission expected from the resolved part of the population where we assume the most pessimistic CTAO detection sensitivity target of $8\times10^{-14}$ ph cm$^{-2}$ s$^{-1}$ (for energies $> 1$ TeV) reported in \cite{CTAConsortium:2023tdz}. This maximizes the expected cumulative intensity of the unresolved part of the population shown in the right panel. (\emph{Bottom row}:) Cumulative intensity $S$ of the resilient (\emph{left}) and fragile (\emph{right}) DM sub-halo populations for a realization matching the median conditions of all simulated realizations, respectively. For definiteness, the intensity is computed for a DM mass of $M_{\chi}$ = 1 TeV assuming annihilation into $b\bar{b}$. The annihilation cross section coincides with the mean values to detect at least one sub-halo for the respective population scenario as reported in Fig.~\protect\ref{fig:SL17_population_detection}. The displayed intensity maps reflect the sub-threshold contribution associated with the MW DM sub-halo population.
\label{fig:GPS-model-intensity}}
\end{figure*}

We analyze the synthetic population of PWNe generated for the CTAO consortium publication on the prospects of characterizing source populations with GPS data \cite{CTAConsortium:2023tdz}. As shown by the authors of this work, PWNe make up the largest fraction of TeV gamma-ray sources along the Galactic plane such that restricting ourselves to them does not limit our qualitative comparison. The results are shown in the top panel of Fig.~\ref{fig:GPS-model-intensity}, which divides the synthetic population of PWNe into a detected and sub-threshold part. The detection threshold follows the quantitative estimate reported in \cite{CTAConsortium:2023tdz} of $8\times10^{-14}$ ph cm$^{-2}$ s$^{-1}$ (for energies $> 1$ TeV). On top of the synthetic sources, we include in the left panel all already known TeV-bright sources also considered in \cite{CTAConsortium:2023tdz}. Note that the authors of this work made sure that bright synthetic sources that have a detected analog in the real sky were removed from the catalog to avoid double counting. The cumulative emission of the astrophysical source population follows the spatial binning defined in Fig.~\ref{fig:intsens}. The results clearly show an increase of gamma-ray emission towards the GC and a Galactic latitude of $0^{\circ}$. This observation is a natural consequence of the formation mechanisms of PWNe, which are remnants of massive stars and supernova explosions at the end of their lifecycle. Thus, their location in the sky is ultimately tied to regions rich in gas and featuring active star formation, for instance, the GC and the MW's spiral arms.

In contrast to most astrophysical source populations, DM sub-halos are not predominantly present along the Galactic plane but may occur throughout the entire Galaxy. In particular, the three-dimensional distribution of the subhalos follows the NFW profile of the Milky Way's main DM halo assumed for the simulations. We report the cumulative gamma-ray emission arising from DM pair-annihilation into $b\bar{b}$ final states in the bottom row of Fig.~\ref{fig:GPS-model-intensity}, which shows one realization of the resilient (left) and fragile (right) sub-halo scenarios. In accordance with the parameter choices adopted in Fig.~\ref{fig:SL17_population_detection}, we fix the DM mass to $M_{\chi} = 1$ TeV and calculate the gamma-ray intensity for a cross section of $\langle\sigma v\rangle = 3.3^{+1.7}_{-1.9}\times10^{-23}$ ($9.3^{+4.4}_{-4.8}\times10^{-23}$) cm$^3$ s$^{-1}$ for the resilient (fragile) population. The stated uncertainties are at the $1\sigma$ level. These cross section values quantify the required annihilation strength to guarantee the detection of at least one DM sub-halo. Thus, the reported cumulative intensities are quantifying the truly diffuse part of the DM sub-halo population. Both realizations illustrate the formerly mentioned spatial distribution of sub-halos. However, the fragile DM sub-halo population is less represented towards the central Galaxy reflecting the fact that the increasing tidal forces deplete this part of the MW from ``fragile'' DM sub-halos.

These figures demonstrate qualitatively and quantitatively that the diffuse emission from the full DM sub-halo population can reach values comparable to or even higher than the diffuse emission expected from the presence of astrophysical gamma-ray emitters. In particular at latitudes above $|b|>1^{\circ}$ DM sub-halos might dominate the diffuse emission in the resilient case and to a lesser extent in the fragile scenario. For example, in the region $1^{\circ}<b<5^{\circ}$ and $85^{\circ}<\ell<265^{\circ}$ the cumulative emission from the resilient population shown in the left panel of Fig.\ref{fig:GPS-model-intensity} reaches a value of $1.2\cdot10^{-11}$ erg/cm$^2$/s which is about five times larger than the expected cumulative flux of the sub-threshold part of the synthetic astrophysical source population. Seen from a different point of view, this implies that these latitudes are less contaminated by astrophysical sources and, hence, they bear a higher potential to detect the brightest part of the DM sub-halo population (which is not featured in both figures). To re-state our earlier results, the brightest sub-halo of the population may be detected at an annihilation cross section about an order of magnitude lower than the cross section applied in this figure. As a side note, these figures show that the presence of astrophysical sources and the thereby caused obstacle of source confusion does not heavily deteriorate the potential to detect individual objects of the population slightly above or below the plane.   

\section{GPS sensitivity to the main dark matter halo}
\label{sec:main-halo}

Since the GPS sweeps the Galaxy all along the Galactic plane, it can potentially detect DM pair-annihilation in the MW's main DM halo. In fact, this signal might even be stronger than any emission from its substructure. We examine it in a simplified setup: We perform a simultaneous fit of the GPS region of interest without the innermost part ($l\in\left[-12^{\circ}, 12^{\circ}\right]$) around the GC. We exclude the central part because it would require special treatment due to the complex astrophysical emission and rather uncertain DM density profile (see \cite{CTA:2020qlo} for details). Next, we assume CR and IE as the only background components. We model the MW parent halo with an NFW profile and parameters taken from the Bayesian characterization in \cite{2017MNRAS.465...76M} where we adopt a realization compatible with the reported posteriors' centers: $r_s=20.5$ kpc, $r_{\odot} = 8.3$ kpc and $\rho_{\odot} = 0.41$ GeV/cm$^3$. While the background components' normalization may vary per energy bin, we assign a single parameter to the DM template that is directly related to the strength of the annihilation cross section. We derive the $5\sigma$ detection sensitivity and associated cross section in the framework discussed in Sec.~\ref{sec:statistics}.

\begin{figure*}[t]
\begin{centering}
\includegraphics[width=0.99\linewidth]{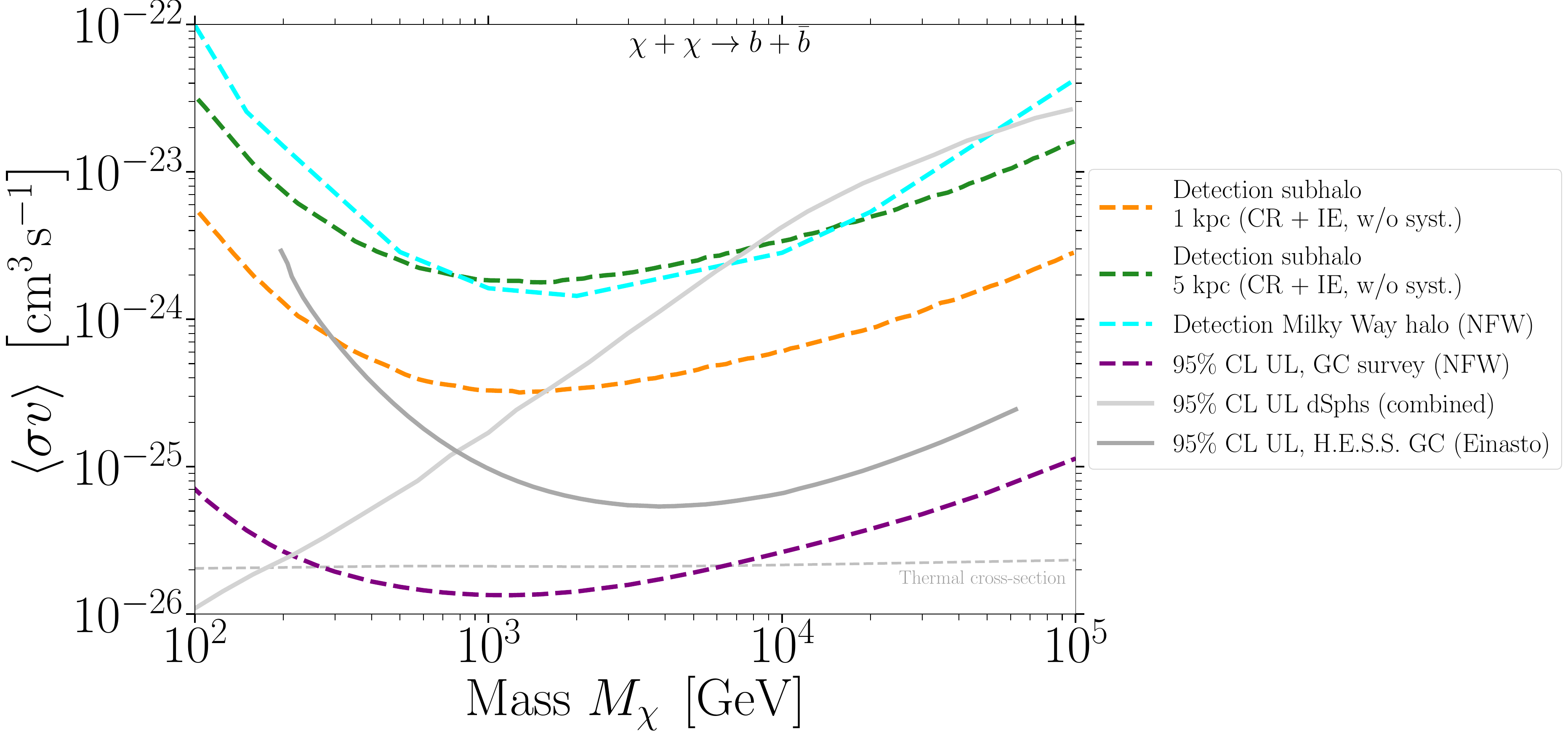}
\par\end{centering}
\caption{Sensitivity to the MW DM parent halo modeled with an NFW profile and representative parameters ($r_s=20.5$ kpc, $r_{\odot} = 8.3$ kpc and $\rho_{\odot} = 0.41$ GeV/cm$^3$) adopted from \cite{2017MNRAS.465...76M} derived in a simultaneous fit to simulated GPS data excluding the region $l\in\left[-12^{\circ}, 12^{\circ}\right]$. The GPS mock data is comprised of the irreducible CR background and interstellar emission. No systematic uncertainties are included.
In addition, the prospects of CTAO's GC survey to constrain the Galactic halo (same NFW profile setup) are stated as 95\% CL UL as dotted purple line \cite{CTA:2020qlo}.
For comparison, we plot the most optimistic prospects for detecting the brightest DM sub-halo of the MW population within the GPS for two distances taken from Fig.~\protect\ref{fig:sigmavM}.
In addition, we show in light gray the 95\% CL UL on the DM annihilation cross section obtained from a combined analysis of 45 dSphs observed by \textit{Fermi}-LAT, HAWC, H.E.S.S., MAGIC and VERITAS \cite{Hess:2021cdp, Kerszberg:2023cup}. The current 95\% CL UL including instrumental systematic uncertainties on DM annihilation in the GC, assuming an Einasto profile, from more than 500h of observation time with the H.E.S.S.~telescope are shown as dark gray line \cite{HESS:2022ygk}. 
\label{fig:MW_mainhalo}}
\end{figure*}

The result for $b\bar{b}$ final states is displayed in cyan as a function of the DM mass $M_{\chi}$ in Fig.~\ref{fig:MW_mainhalo}. Additionally, we plot the results for the brightest DM sub-halo from Fig.~\ref{fig:sigmavM} for direct comparison. As expected, the MW parent halo constitutes a bright extended gamma-ray source in the GPS. Depending on the position of the brightest DM sub-halo, the main halo may even be the most prominent target in the GPS. However, our analysis is meant to show a rather simplified scenario with only two background components and no further complications. In reality, it is necessary to include all resolved, localized gamma-ray emitters in the GPS (or to mask them) in order to search for such a substantially extended component. Realistically speaking, detecting the MW parent halo would require even larger annihilation cross section.

We stress, however, that the GC and the planned GC survey proposed by the CTAO consortium have a much better sensitivity to the annihilation signal of the main halo as indicated by the purple line in Fig.~\ref{fig:MW_mainhalo} taken from \cite{CTA:2020qlo}. The quoted sensitivity is stated in terms of a 95\% confidence level (CL) upper limit (UL) for the same NFW profile -- in contrast to the $5\sigma$ discovery reach presented in this work. Yet, the DM density profile towards the GC is rather unconstrained as concerns the results of parameter estimation from data on stellar kinematics \cite{Karukes:2019jxv, Benito:2020lgu}. Consequently, DM density profiles with extensive cores in the MW's center are not excluded and would drastically weaken the sensitivity of CTAO by orders of magnitude (cf.~\cite{CTA:2020qlo}).

\section{Conclusions}
\label{sec:conclusions}

In this work, we examined the prospects of CTAO's GPS to characterize the substructure of the MW's parent DM halo. The presence of sub-halos within the MW is a natural consequence of the cold DM paradigm. Yet, the number and distribution of DM sub-halos are subject to tidal forces acting on them due to the MW's and its components' gravitational potential. Hence, we investigated these objects relying on two benchmark scenarios quantifying the survival probability of DM sub-halos inside the Galactic halo (see \cite{Stref:2016uzb}); an optimistic case -- the resilient population -- which withstands stronger tidal forces and thereby evades total disruption to a larger extent than sub-halos of the more pessimistic scenario -- the fragile population. Beyond these scenarios for DM substructures other formation mechanisms and models have been proposed that yield more aggressive predictions in number and density, for example, in the series of works \cite{White:2022yoc, Delos:2022bhp, Delos:2023ipo, Delos:2023azx}. Here, highly concentrated and low-mass cusps may form in the early universe that are very different from the typical DM sub-halos predicted by $N$-body simulations of structure formation. 

Assuming that the particle nature of DM is a form of WIMP, DM sub-halos may efficiently produce (very) high-energy gamma-ray emission due to DM pair-annihilation in their innermost region boosted by the expected high DM densities. Depending on the distance to the observer and the sub-halo's mass, sub-halos could appear as point-like or extended gamma-ray sources enriching the zoology of known gamma-ray emitters. Gamma-ray telescopes like CTAO with their high sensitivity and good angular resolution are thus suitable to detect a signal from these objects. Previous sensitivities forecasts have focused on the CTAO consortium's proposed extragalactic survey (covering about one-quarter of the sky) as well as serendipitous signals in the entirety of CTAO's legacy data \cite{Hutten:2016jko, CoronadoBlazquez:2021}. Our work investigates the potential of the GPS -- a large-scale survey of the CTAO consortium not yet considered in the context of searches for DM sub-halos -- to detect these objects. While the extragalactic survey is less ``polluted'' with conventional astrophysical gamma-ray emitters and comprises a more considerable fraction of the sky, thus facilitating a DM sub-halo detection, the GPS is characterized by a larger exposure time, which improves the sensitivity to point-like and slightly extended sources. Concerning the brightest sub-halo of the population, we find that GPS prospects may be about a factor of three better than similar considerations for the extragalactic survey \cite{Hutten:2016jko} depending on the position of the sub-halo. 


We investigated the impact of several sources of systematic uncertainty that may be of relevance to our inferred $5\sigma$ discovery reach of the brightest sub-halo of $\langle\sigma v\rangle \sim \mathcal{O}(10^{-25} - 10^{-24})$ cm$^3$/s. We recall from our discussions above that such cross-section values, despite being certainly larger than the canonical thermal annihilation cross section, can be sufficient to generate the observed DM abundance in the universe if effects like Sommerfeld enhancement or resonances are present. Additionally, our results can potentially be relaxed when accounting for the additional boost of the sub-halos' $J$-factor due to sub-substructure.

Based on the sub-halo extensions and $\mathcal{J}$-factors derived from many realizations of the assumed population scenarios, we find that the prospect of detecting at least one member of the population requires $\langle\sigma v\rangle \sim \mathcal{O}(10^{-23})$ cm$^3$/s (see Sec.~\ref{sec:pop-study}).  

As reported in Sec.~\ref{sec:sens_iemsyst}, diffuse interstellar emission on top of the irreducible CR background weakens our prospect at the level of tens of per cent in case of perfect knowledge about its spatial morphology. Imperfect knowledge -- a scenario encountered at the GeV scale with \textit{Fermi}-LAT -- introduces uncertainties as large as a factor of two around TeV energies (see \cite{Eckner:2022gil}). Considering the effect of instrumental systematic uncertainties of around $3\%$ (CTAO's instrumentation goal) yields a deterioration of the expected sensitivity around a factor of two. More pessimistic assumptions around $10\%$ increase detectable annihilation cross section by an order of magnitude. 

We found in Sec.~\ref{sec:source-discrimination} that a once-detected DM sub-halo is readily distinguished from a simple point-like source or an extended source modeled with a Gaussian profile. This highlights the importance of the very distinct sub-halo profile to evade the negative impact of source confusion in crowded regions towards the Galactic equator. We also motivated in Fig.~\ref{fig:GPS-model-intensity} that the search for DM sub-halos slightly above and below the Galactic equator is less impacted by conventional astrophysical sources. 


Current generation space-borne and ground-based gamma-ray instruments are already somewhat constraining the viable parameter space of TeV-scale DM.
We illustrate the current landscape of 95\% CL upper limits on $\langle\sigma v\rangle$ in Fig.~\ref{fig:MW_mainhalo}, which includes results from the combined observations of Galactic dSphs with \textit{Fermi}-LAT, H.E.S.S., MAGIC, VERITAS and HAWC \cite{Hess:2021cdp, Kerszberg:2023cup} and H.E.S.S.~observations of the GC for more than 500h \cite{HESS:2022ygk}. All these observations, of course, come with their own assumptions, data selection, quality cuts, and systematic uncertainties. In particular, the very stringent H.E.S.S.~limits from the GC are subject to large uncertainties of the MW's innermost DM profile. Currently, a peaked profile like the shown Einasto distribution is as likely as a more extended/cored profile which heavily deteriorates the sensitivity of H.E.S.S.~to a DM signal in the GC by orders of magnitude. Therefore, the combined limits from dSphs are the more reliable observational constraint. In light of these, the brightest part of the sub-halo population may provide complementary insight into the WIMP parameter space should it fall into the CTAO's GPS band. Thus, CTAO's GPS has sufficiently good sensitivity to search for the dark substructure of the MW's main halo and to set bounds on the annihilation cross section in case of no detections. 

\begin{acknowledgments}
We warmly thank the CTAO Consortium internal referees Nagisa Hiroshima, Daniel Kerszberg and Anna Lipniacka as well as Torsten Bringmann for your careful reading of the manuscript and helpful comments to improve its clarity and delivery.
The work of CE is supported by the ANR through grant ANR-19-CE31-0005-01 (PI: F.~Calore), and has been supported by the EOSC Future project which is co-funded by the European Union Horizon Programme call INFRAEOSC-03-2020, Grant Agreement 101017536. This publication is supported by the European Union's Horizon Europe research and innovation programme under the Marie Sk\l odowska-Curie Postdoctoral Fellowship Programme, SMASH co-funded under the grant agreement No.~101081355. Research at Perimeter Institute is supported in part by the Government of Canada
through the Department of Innovation, Science and Economic Development Canada and by the Province of Ontario through the Ministry of Colleges and Universities. \\ This work was conducted in the context of the CTAO `Dark Matter and Exotic Physics' Working Group.
\end{acknowledgments}

\bibliographystyle{bibi}
\bibliography{bibliography}

\appendix

\end{document}